\documentclass[10pt,floats,letterpaper]{article}
\usepackage{amsfonts,color,epsfig}
\renewcommand{\theequation}{A-\arabic{equation}}
\newcommand{\nc}{\newcommand}

\newcommand{\rnc}{\renewcommand}

\makeatletter
\rnc{\theequation}{\thesection.\arabic{equation}}
\@addtoreset{equation}{section}
\makeatother
\nc{\fig}[5]{
\begin{figure}[!htbp]
    \begin{center}
    \leavevmode
    \centerline{
        \includegraphics[width=#1, height=#2]{#3}
        }
    \caption[]{#4}
    \label{#5}
    \end{center}
\end{figure}}
\nc{\figs}[8]{
\begin{figure}[!htbp]
    \begin{center}
    \leavevmode
    \centerline{
        \includegraphics[width=#1, height=#2]{#3}
        \includegraphics[width=#4, height=#5]{#6}
        }
    \caption[]{#7}
    \label{#8}
    \end{center}
\end{figure}}
\begin{document}
\begin{flushright}
{\tt gr-qc/0609094}
\end{flushright}
\vspace{2mm}
\begin{center}
{{{\Large \textbf{Gravitationally Collapsing Shells in (2+1) Dimensions }}}}\\[16mm]
{Robert B. Mann\footnote{%
mann@avatar.uwaterloo.ca} and John J. Oh\footnote{%
j4oh@sciborg.uwaterloo.ca}}\\[10mm]
{Department of Physics, University of Waterloo,\\[0pt]
Waterloo, Ontario, N2L 3G1, Canada}\\[0pt]
\end{center}
\vspace{8mm}
\begin{abstract}
We study gravitationally collapsing models of pressureless dust,
fluids with pressure, and the generalized Chaplygin gas (GCG) shell
in (2+1)-dimensional spacetimes. Various collapse scenarios are
investigated under a variety of the background configurations such
as anti-de Sitter(AdS) black hole, de Sitter (dS) space, flat and
AdS space with a conical deficit. As with the case of a disk
of dust, we find that the collapse of a dust shell coincides with
the Oppenheimer-Snyder type collapse to a black hole provided the
initial density is sufficiently large.  We also find -- for all
types of shell -- that collapse to a naked singularity is possible
under a broad variety of initial conditions. For shells with
pressure this singularity can occur for a finite radius of the
shell. We also find that GCG shells exhibit diverse collapse
scenarios, which can be easily demonstrated by an effective
potential analysis.
\end{abstract}
\vspace{10mm}


\vspace{3cm}

\hspace{11.5cm}{Typeset Using \LaTeX}
\newpage
\section{Introduction}
\label{sec:intro}

Over the last few decades, general relativity in (2+1) dimensions has
fascinated both field theorists and relativists because of its fertility as
a test-bed for ideas about quantum gravity. One particular feature of interest
is manifest when a negative cosmological constant is present.  Despite
the fact that the spacetime geometry of this solution is an
anti-de Sitter (AdS) spacetime, possessing negative
constant curvature, a black hole can be present under a suitable
choice of topological identifications \cite{btz}. This solution has drawn much attention
since its inception from a wide variety of perspectives \cite{3dgravity}.

Shortly after the black hole solution was obtained, it was shown that
it can be formed from a disk of pressureless dust undergoing gravitational collapse \cite{mr}
(the three-dimensional analogue of Oppenheimer-Snyder type collapse),
generalizing earlier results that suggested matter could collapse to form conical
singularities \cite{gak}.
Further study on this subject has been carried out from several viewpoints,
including the formation of a black hole from colliding point particles \cite{mat}
and the more recent demonstration of critical phenomena in the context
of collapse \cite{3dcollapse}.
These results are consistent with other results in four dimensions as well as results in
two dimensions \cite{mr2}.

Recently, a cosmological model of a (generalized) Chaplygin gas (GCG) was
introduced as a possibile explanation of the present acceleration of the
universe, the existence of dark energy, and the unification of dark
energy and dark matter \cite{kmp,btv,bbs}.
Historically its original motivation was to account for the lifting force on
a plane wing in aerodynamics \cite{chap}. Afterwards, the same equation of
state was rediscovered in the context of aerodynamics \cite{tsien,karman}. A
more interesting feature of this gas was recently renewed in an intriguing
connection with string theory, insofar as its equation of state can be obtained
from the Nambu-Goto action for $d$-branes moving in a $(d+2)$-dimensional
spacetime in the light-cone frame \cite{bh}. In addition, it has been shown
that the Chaplygin gas is, to date, the only fluid that admits a
supersymmetric generalization \cite{hjp}; the relevant symmetry group was
described in ref. \cite{bj}. Moreover, further theoretical developments of
the GCG were given in terms of cosmology and astrophysics \cite{coas}.
Inspired by the fact that the Chaplygin gas has a negative pressure,
violating the energy conditions (in particular the null energy condition
(NEC)), traversable wormhole solutions were found in four dimensions \cite%
{lobo}.

It is natural to ask whether or not a black hole can be formed from
gravitational collapse of this gas in a finite collapse time. Much
of the work on black hole formation deals with pressureless dust
collapse; collapse of  this kind of exotic fluid to black holes so
far has not received much treatment. Recent work \cite{dc} \
involved investigation of spherically symmetric clouds of a
collapsing modified Chaplygin gas in four dimensions, where it was
shown that it always leads to the formation of a black hole.

In this paper, we investigate some gravitational collapse scenarios
of shells with a variety of equations of state, including the GCG
shell. To set the stage we first consider the collapse of a shell of
pressureless dust. In  dust collapse scenarios the evolution of the
system is obtained by matching the inside and outside geometries
using the junction conditions \cite{isr,chase,poi},
\begin{equation}
  \label{eq:junctions}
  [g_{ij}]=0,~~[K_{ij}]=0,
\end{equation}
where $[h] \equiv h_{+} - h_{-}$ and ($+$) and ($-$) represent
exterior and interior spacetimes, respectively. However for shells
with pressure the junction condition for the extrinsic curvature in
eq. (\ref{eq:junctions}) is no longer valid, since there is a
nonvanishing surface stress-energy on the boundary of the shell to
take into account.

The main result of our investigation is that gravitational collapse
in (2+1) dimensions does not necessarily lead to black hole
formation for any of the fluid sources we study. The end points of
collapse depend on the initial conditions, and can lead to either a
black hole or the formation of a singularity and a Cauchy
horizon\footnote{This latter point has been overlooked in previous
studies \cite{gak}.}. This singularity is characterized by the onset
of a divergent stress energy in the shell, whose intrinsic Ricci
scalar also diverges in finite proper time for observers comoving
with the shell.  For pressureless dust the singularity develops when
the shell collapses to zero size. However for shells with pressure
the singularity develops at some nonzero size characterized by the
equation of state.  A similar scenario holds for the GCG shell. We
also find that collapse is not the only possibility, but that shells
can also expand out to infinity, possibly with a bounce depending on
the initial conditions. Our results are consistent with earlier work
on shell collapse in (2+1) dimensions \cite{Olea,Steif},
generalizing them to include a more detailed analysis of collapse to
naked singularities, and to situations in which a more general
relationship between density and pressure is assumed.

The outline of our paper is as follows.
In section \ref{sec:shell}, we briefly present a
formulation of the shell collapse and obtain the evolution equation
for the dust shell radius. In section \ref{sec:dustshell}, the gravitational
collapses of pressureless dust shell are studied and compared to the
result of dust cloud collapse in \cite{mr}.
In section \ref{sec:pressure}, we study a collapse of a shell with an arbitrary pressure with no loss of generality.
In section \ref{sec:GCG}, the collapse of GCG shell is studied and
some possible collapse conditions are found. Finally, we shall summarize and discuss our results in
section \ref{sec:discussions}. We consider the construction of some relevant Penrose diagrams and some
basic properties of Jacobian elliptic functions in appendices.

\section{Shell Collapse }
\label{sec:shell}

We assume that the metrics in both regions, ${\mathcal V}_{+}$
(outside the shell) and
${\mathcal V}_{-}$ (inside the shell) are given by
\begin{equation}
  \label{eq:1}
  (ds)^2_{{\mathcal V}_{\pm}} = -F_{\pm} dT^2 + \frac{dR^2}{F_{\pm}} + R^2d\theta^2,
\end{equation}
where $F_{+}$ and $F_{-}$ are exterior and interior metrics, respectively. The surface stress-energy
for a fluid of density $\rho$ and pressure $p$ is
\begin{equation}
S^{ab} = \rho u^{a}u^{b} + p h^{ab} ,\label{eq:1.5}
\end{equation}
where $h_{ab}=g_{ab}+u_a u_b$ is an induced metric on $\Sigma$, and $u^{a}$ is the
shell's velocity. For  dust $p=0$, whereas $p=-A/\rho^{\alpha}$ for the generalized Chaplygin gas (GCG).
We employ a coordinate system   ($\tau$, $\theta$) on $\Sigma$;  at $R={\mathcal R}(\tau)$ the induced
metric is
\begin{eqnarray}
  \label{eq:2}
  (ds)_{\Sigma}^2 &=& - F_{\pm} dT^2 + \frac{d{\mathcal R}^2}{F_{\pm}} +
  {\mathcal R}(\tau)^2 d\theta^2\nonumber\\
  &=& - d\tau^2 + r_{0}^2 a^2(\tau)d\theta^2,
\end{eqnarray}
Continuity of the metric implies that $[g_{ij}]=0$ or
 $F_{\pm}^2 (d{T}/d\tau)^2 = (d{\mathcal R}/d\tau)^2 + F_{\pm}$
and ${\mathcal R}^2 = r_{0}^2 a(\tau)^2$ . However there exists a discontinuity in
the extrinsic curvature of the shell,
$[K_{ij}]\ne 0$, since  nonvanishing surface stress-energy exists.
The extrinsic curvatures on ${\mathcal V}_{\pm}$ are
\begin{equation}
  \label{eq:3}
  K_{\tau\tau}^{\pm} = -\frac{d}{d\mathcal R} \sqrt{\left(\frac{d{{\mathcal
  R}}}{d\tau}\right)^2 + F_{\pm}}, ~~K_{\theta\theta}^{\pm} =
  \frac{1}{\mathcal R} \sqrt{\left(\frac{d{\mathcal R}}{d\tau}\right)^2 + F_{\pm}}.
\end{equation}
The surface stress-energy is defined by
\begin{equation}
  \label{eq:4}
  {\mathcal S}_{ab} = - \left([K_{ab}]-[K]h_{ab}\right),
\end{equation}
where $h_{ab}$ is the induced metric on $\Sigma$.
On the edge of the shell
\begin{equation}
  \label{eq:einbein}
  e^{\alpha}_{(\tau)} = \left(\frac{dT}{d\tau}, \frac{d{\mathcal
  R}}{d\tau}, 0\right),~~e^{\alpha}_{(\theta)} =
  \left(0,0,\frac{1}{\mathcal R}\right),
\end{equation}
and
\begin{equation}
  \label{eq:normal}
  n_{\alpha} = \left(-\frac{d{\mathcal R}}{d\tau}, \frac{dT}{d\tau}, 0\right),
\end{equation}
where $\alpha$ is a coordinate system in the bulk, $(T,R,\theta)$. The surface stress-energy can be straightforwardly evaluated
\begin{equation}
  \label{eq:5}
  S^{\tau}_{~\tau} = \frac{1}{{\mathcal R}}
  (\beta_{+}-\beta_{-}),~~S^{\theta}_{~\theta} = \frac{d}{
  d{\mathcal R}}({\beta}_{+} - {\beta}_{-}),
\end{equation}
where $\beta_{\pm} \equiv \sqrt{(d{\mathcal R}/d\tau)^2 + F_{\pm}}$. Using
eqs. (\ref{eq:1.5}) and (\ref{eq:5}), we have two relations,
\begin{eqnarray}
  -\rho &=& \frac{1}{{\mathcal R}}
  (\beta_{+}-\beta_{-}) \label{eq:6}\\
  p &=& \frac{d}{
  d{\mathcal R}}({\beta}_{+} - {\beta}_{-}).\label{eq:7}
\end{eqnarray}

The preceding relations can be written in the form
\begin{eqnarray}
 \beta_{+}-\beta_{-} +\rho {\mathcal R} &=&0 \label{eq:6a}\\
\frac{d}{d{\mathcal R}}\left(\rho {\mathcal R}\right) +  p &=& 0
\end{eqnarray}
and eq.(\ref{eq:6a}) implies for positive densities that
$\beta_+ <\beta_-$, which in turn implies that
\begin{equation}\label{poscond}
a_{-}{\mathcal R}^2/\ell^2 + k_{-} > a_+{\mathcal R}^2/\ell^2 + k_+
\end{equation}
where the generic form of the metrics we study have $F_{+}=(a_+{\mathcal
R}^2/\ell^2 + k_+)$ and $F_{-}= (a_{-}{\mathcal R}^2/\ell^2
+k_{-})$. Here $a_{\pm}$ and $k_{\pm}$ are constants whose values
respectively determine whether or not the spacetime is
asymptotically AdS, dS, or flat, and whether or not the spacetime
contains a point mass or a black hole. Differing magnitudes for
$a_{\pm}$ correspond to different values of the size of the
cosmological constant inside and outside of the shell. Without loss
of generality we can choose one of these to have unit magnitude,
i.e. $|a_{-}|=1$, though we shall not always exercise this option.

\section{Collapse of a Pressureless Dust Shell}
\label{sec:dustshell}

For the dust shell case, $p=0$ and eq. (\ref{eq:7}) becomes
\begin{equation}
  \label{eq:12}
  0 = \frac{d}{d{\mathcal R}} (\beta_{+}-\beta_{-}).
\end{equation}
Eq. (\ref{eq:12}) is easily solved; equating with eq. (\ref{eq:6})
yields
\begin{equation}
  \label{eq:13}
  (\beta_{+}-\beta_{-}) = - \rho_{0} = -\rho {\mathcal R},
\end{equation}
where $\rho_{0}$ is an integration constant. The density profile is therefore
$\rho=\rho_{0}/{\mathcal R}$; if $\rho=0$ then clearly $F_{+}=F_{-}$.

Eq. (\ref{eq:13}) yields the generic differential equation for the
dust shell
\begin{equation}
  \label{eq:14b}
  \sqrt{\left(\frac{d{\mathcal R}}{d\tau}\right)^2 + \frac{a_{+}{\mathcal R}^2}{\ell^2} +k_+} -
  \sqrt{\left(\frac{d{\mathcal R}}{d\tau}\right)^2
  + \frac{a_{-}{\mathcal R}^2}{\ell^2} +k_{-}} + \rho_{0} = 0.
\end{equation}
Upon redefining $t \equiv \tau/\ell$, $x \equiv {\mathcal R}/\ell$,
$\varrho \equiv \rho\ell$, and $\varrho_{0} \equiv \rho_{0}\ell$,
 eq. (\ref{eq:14b}) can be written as
\begin{equation}
  \label{eq:re14b}
  \sqrt{\dot{x}^2 + a_{+}x^2 + k_{+}} - \sqrt{\dot{x}^2 + a_{-}x^2 + k_{-}} + \varrho_{0} = 0,
\end{equation}
which can be alternatively written in the form
\begin{equation}
\label{eq:eqdusteffp}
\dot{x}^2 + V_{\rm eff}(x) = 0,
\end{equation}
where the effective potential is given by
\begin{equation}
\label{eq:effp}
V_{\rm eff}(x) = -\frac{1}{4\varrho_{0}^2}(a_4 x^4-2a_2 x^2+a_0)
\end{equation}
with
\begin{equation}
  \label{eq:sol14bconst2}
a_4=(a_{+}-a_{-})^2 \qquad
a_2=(a_{+}+a_{-})\varrho^2_{0}-(a_{+}-a_{-})(k_{+}-k_{-}) \qquad
a_0=\left(\varrho^2_{0}-(k_{+}+k_{-})\right)^2-4k_{+}k_{-}.
\end{equation}
Equation (\ref{eq:re14b}) has the general solution
\begin{equation}
  \label{eq:sol14b}
 x(t)=-\frac{\gamma_{-}}{|a_{+} - a_{-}|}
 {\rm JacobiSN}\left[\frac{\gamma_{+}}{2\varrho_{0}}(t-t_{0}), {\frac{\gamma_{-}}{\gamma_{+}}}\right]
\end{equation}
where
\begin{equation}
  \label{eq:sol14bconsts}
\gamma^2_{\pm}=a_{2}\pm\sqrt{a^2_2-a_0a_4}
\end{equation}
and
\begin{equation}
\label{eq:kappas} t_0 = \frac{2\varrho_{0}}{\gamma_{+}}{\rm
JacobiSN}^{-1}\left[\frac{|a_+-a_-|}{\gamma_{-}}x_{0},
\frac{\gamma_-}{\gamma_+}\right] =
\frac{2\varrho_{0}}{\gamma_{+}}{\rm
EllipticF}\left[\frac{|a_+-a_-|}{\gamma_{-}}x_{0},
\frac{\gamma_-}{\gamma_+}\right]  .
\end{equation}
The properties of the Jacobi elliptic functions are reviewed in an appendix.

In the special case that $a_+=a_-$, $\gamma_-=0$ and the solution
becomes
\begin{equation}
  \label{eq:sol14bb}
 x(t)= -\sqrt{\frac{a_0}{2a_2}} \sin\left[\frac{\sqrt{2a_2}}{2\varrho_{0}}(t-t_{0})\right].
\end{equation}
Alternatively, if $a_0=0$, then the solution is
\begin{equation}
  \label{eq:sol14bc}
 x(t)= -
 \sqrt{\frac{2a_2}{a_4}}\frac{1}{\sin\left[\frac{\sqrt{2a_2}}{2\varrho_{0}}(t-t_{0})\right]}.
\end{equation}

The qualitative behaviour of the solutions will depend upon the
relative signs of the four parameters $a_{\pm}$ and $k_{\pm}$. In
general there are 81 possibilities since each parameter can vanish,
though of course not all of these are allowed.  For example if
$a_+<0$ then $k_{+}>0$ in order to preserve the metric signature.
There are additional restrictions that arise from the reality of the
collapse trajectory, which imply that the quantities $\gamma_{\pm}$
must either be pure real or pure imaginary. This yields
$a^2_2-a_0a_4>0$, or
\begin{equation}
  \label{eq:reality}
a_{+}a_{-}\varrho^2_0 + (a_{+} - a_{-})(a_{+}k_{-} - a_{-}k_{+}) =
a_{+}a_{-}(\varrho^2_0 - k_+ - k_-) + a^2_{+}k_{-} + a^2_{-}k_{+}
> 0.
\end{equation}

Much of the general behaviour of the solution can be
understood by noting that eq. (\ref{eq:eqdusteffp}) describes the
one-dimensional motion of a point particle of zero energy in the
effective potential $V_{\rm eff}$ given in eq. (\ref{eq:effp}), which is
sketched in Fig. \ref{fig:effectptdust}.  Note that only
the $x>0$ part of the potential is relevant; the behaviour of the
shell will depend on the number of roots of the effective potential
in this region.

\fig{10.5cm}{9.5cm}{effectptdustcol}
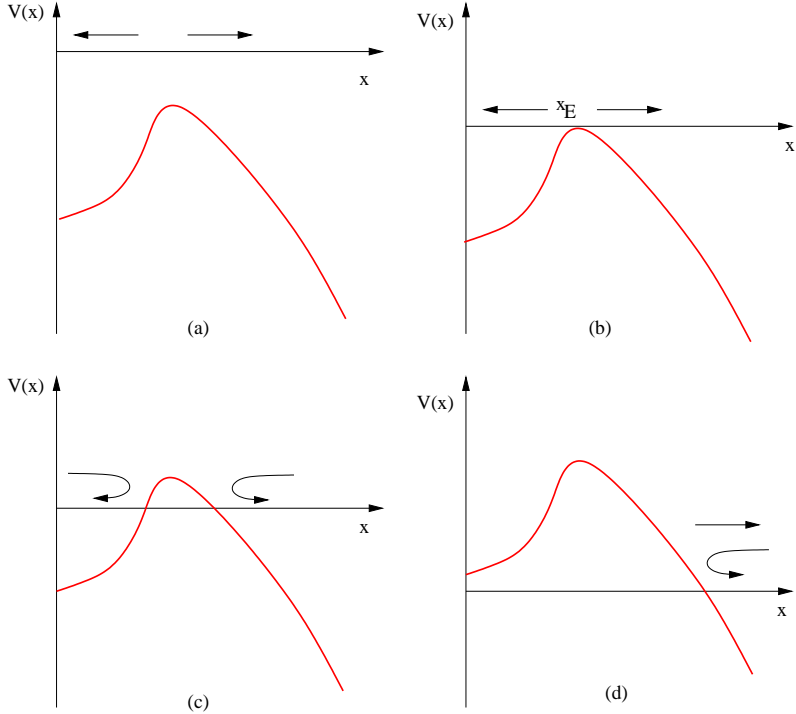
{\small Plots of the effective potential of the dust shell for existing roots.}
{fig:effectptdust}

If there are no roots, then the shell will either collapse to zero
size from some finite value, or it will expand indefinitely,
depending upon the initial conditions. If there is one
non-degenerate root then the shell will either expand indefinitely
or contract to some finite size and then expand (for example eq.
(\ref{eq:sol14bc}) describes this situation). If there is one
degenerate root, then the shell can either (a) sit in an unstable
equilibrium at some fixed value $x=x_E=\sqrt{a_2/a_4}$ (provided
$a^2_2=a_0a_4$), (b) collapse to either a black hole or a naked
singularity provided its initial size is such that $x_0<x_E$, or (c)
exhibit the behaviour of the non-degenerate single root case,
provided $x_0
> x_E$. If there are two roots, then there will either be collapse
to a black hole or naked singularity, or else the behaviour will be
qualitatively similar to that of the non-degenerate single root
case. These various cases are illustrated in Figure
\ref{fig:effectptdust} by the arrows that indicate possible
trajectories of the shell and resemble in part the
higher-dimensional situation \cite{Alberghi,MyHub}.The key
distinction here is the possible of collapse of the shell either to
zero size or to a black hole, depending on the choice of parameters
and the initial conditions.
 A discussion of the Penrose
diagrams for a number of these scenarios appears in the appendix.

The preceding analysis assumed $a_4\ne 0$. If $a_4 = 0$ (ie.
$a_+=a_-$), then the effective potential is quadratic. If $a_+>0$
the shell will always collapse provided $a_0>0$, whereas if  $a_+<0$
then the shell will either expand indefinitely or contract to some
finite size and then expand, or -- if $a_0<0$ -- collapse to a naked
singularity.

Note that collapse to a point mass is not a possible end
state for the dust shell. One gets a hint of the underlying problem
upon realizing that $\dot{x}$ will not be zero at the end point of
collapse. This suggests a bounce, but since the interior spacetime
has shrunk to zero size, the future evolution of the spacetime after
such a putative bounce is not uniquely determined.  Instead, as the shell collapses
its induced curvature becomes singular as $t\to t_0$, as is clear
from the expansion of the Ricci scalar associated with the induced
metric (\ref{eq:2})
\begin{equation}
  \label{eq:Ric-exp}
 R^\mu_\mu[\Sigma] =\frac{1}{x^2(t)}\left(x(t) \frac{d^2 x}{dt^2} -
 \left(\frac{dx}{dt}\right)^2\right)= -\frac{1}{2(t-t_0)^2} + \cdots
\end{equation}
which diverges quadratically regardless of the values of the
parameters $a_{\pm}$ and $k_{\pm}$.

If $k_+<0$ this singularity will be cloaked by an event
horizon. However if $k_+ > 0$ the ostensible point mass end state
suggested by the form of the exterior metric  will actually be an incomplete spacetime,
with a Cauchy horizon emerging from the singularity.
There is nothing {\it a-priori} to prevent this choice
for $k_+$, and so there will be a range of initial conditions (even
for asymptotically flat space) in which the end state of collapse
yields a naked singularity, in violation of cosmic censorship.

We shall now consider the evolution of the shell in more specific
terms, categorizing our study by $a_+>0$ (exterior AdS space),
$a_+=0$ (exterior flat space) and $a_+<0$ (exterior dS space).

\subsection{External AdS Space}

If the spacetime is asymptotically anti-de Sitter, then
$a_+>0.$  The general solution is then given by (\ref{eq:sol14b})
\begin{equation}
  \label{eq:sol14b-AdS}
 x(t)=-\frac{\gamma_{-}}{|1 - a_{-}|}
 {\rm JacobiSN}\left[\frac{\gamma_{+}}{2\varrho_{0}}(t-t_{0}),
 {\frac{\gamma_{-}}{\gamma_{+}}}\right]
\end{equation}
where $\gamma_{\pm}$ is still given by eq. (\ref{eq:sol14bconsts}),
but with
\begin{equation}
  \label{eq:sol14bconstAdS}
a_4=(1-a_{-})^2 \qquad a_2=(1+a_{-})\varrho^2_{0}+(1-a_{-})(M+k_{-})
\qquad a_0=\left(\varrho^2_{0}+(M-k_{-})\right)^2 + 4 M k_{-}.
\end{equation}
where we have set $a_+=1$ without loss of generality and $k_+=-M$
so that a naked singularity is avoided.

Collapse to an AdS$_3$ black hole with no angular momentum
and with cosmological constant, $\Lambda=-1/\ell^2$ will take place
provided
\begin{equation}
  \label{eq:adsBHcond}
  x(0)=x_0 <
\frac{\sqrt{(1+a_{-})\varrho^2_{0}+(1-a_{-})(M+k_{-})-2\varrho_{0}\sqrt{a_{-}(\varrho^2_0
+ M - k_-) + k_{-} - a^2_{-}M}}}{|1-a_{-}|}
\end{equation}
and that the term under the second square root is positive, which is
the condition (\ref{eq:reality}). If (\ref{eq:adsBHcond}) is
satisfied, we have an additional condition for the collapse,
$\dot{x}_0^2 + x^2_{0} \ge {M}=x_{H} > -k$ (with $\dot{x_0}$ the
initial velocity of the shell), which means that the initial radius
cannot be smaller than the black hole horizon. If these conditions
are satisfied and $\dot{x}_0>0$ then the shell will first expand to
a maximal size given by the right-hand-side of (\ref{eq:adsBHcond}),
and then collapse to a black hole; otherwise the shell will
irreversibly collapse. If (\ref{eq:adsBHcond}) is violated, then the
shell collapses to a minimal radius
\begin{equation}
  \label{eq:adsexpand}
  x_{min}
=\frac{\sqrt{(1+a_{-})\varrho^2_{0}+(1-a_{-})(M+k_{-})+2\varrho_{0}\sqrt{a_{-}(\varrho^2_0
+ M - k_-) + k_{-} - a^2_{-}M}}}{|1-a_{-}|}
\end{equation}
and then expands to indefinitely large size (or else sits at
a point of unstable equilibrium if $a_{-}(\varrho^2_0 + M - k_-) +
k_{-} - a^2_{-}M =  0$ and if the initial conditions are properly
set).

If $k_{-}=1$, the interior is pure AdS; if $1> k_- >0$, then the
interior metric corresponds to a point mass in AdS spacetime; if
$k_-<0$ then the interior metric corresponds to a black hole.
Since eq. (\ref{poscond}) implies that
\begin{equation}
  \label{eq:rhoclass}
  \varrho_{0} = \left\{\begin{array}{l}
                      {\rm positive},~~((a_- -1)x_0^2+ k_-+M>0) \\
                      {\rm negative},~~((a_- -1)x_0^2+ k_-+M<0) \\
                      0,~~~~~~~~~~~((a_- -1)x_0^2+ k_-+M=0). \\
                    \end{array} \right.
\end{equation}
We see for positive energy density that in general the mass
of the interior black hole must be smaller than $M$ provided  $0 <
a_- < 1$, ie. the magnitude of the interior cosmological constant is
not as large as that of the exterior space. If $a_- \leq 0$ then
$k_->0$ or else the metric signature of the interior is not properly
preserved (alternatively condition (\ref{eq:reality}) is not
satisfied).  If $a_- \geq 1$ then the interior black hole will have
a larger mass than $M$, with the energy of the shell contributing
negatively to the total energy of the spacetime.

The case $a_+=a_-=1$ merits special attention, since it
corresponds to the previously analyzed collapse of a disk of dust
\cite{mr}. We recover the solution (\ref{eq:sol14bb}), which can be
written as
\begin{equation}
  \label{eq:analysol}
 x(t)=-\frac{\sin (t-t_0)}{2\varrho_{0}} \sqrt{M^2+2(\varrho_{0}^2+k) M +
  (\varrho_{0}^2 -k)^2},
\end{equation}
where
\begin{equation}
  \label{eq:inconstdet}
t_0 =  {\rm arcsin}\left[\frac{2 \varrho_{0}
x_{0}}{\sqrt{M^2+2(\varrho_{0}^2+k) M +
  (\varrho_{0}^2 -k)^2}}\right]={\rm arcsin}\left[\frac{ x_{0}}{\sqrt{\dot{x_0}^2+{x_0}^2}}\right].
\end{equation}
Collapse to a black hole takes place when
 $x(t_H)={x}_{H} \equiv {\mathcal R}_{H}/\ell=\sqrt{M}$, where
\begin{equation}
\label{eq:horcoltime}
 t_H =  {\rm arcsin}\left[\frac{ x_{0}}{\sqrt{\dot{x_0}^2+{x_0}^2}}\right]
 -  {\rm arcsin}\left[\frac{\sqrt{M}}{\sqrt{\dot{x_0}^2+{x_0}^2}}\right]
\end{equation}
which is when (in comoving time) the radius of the shell is
coincident with the event horizon.

Note that eq. (\ref{eq:analysol}) yields the requirement that $x(t)$
be real, implying in turn that $a_0>0$ or
\begin{equation}
\label{eq:requires} M^2+2(\varrho_{0}^2+k_-) M +
(\varrho_{0}^2-k_-)^2 =(\varrho_0^2-(k_- + M))^2+ 4\varrho_0^2 M \ge
0
\end{equation}
which is satisfied for all positive $M$. The effective potential is
obtained by setting $a_{+}=a_{-}=1$, $k_{+}=-M$ in eq.
(\ref{eq:effp}),
\begin{equation}
\label{eq:effp1}
V_{\rm eff}(x)=\frac{1}{4\varrho_{0}^2}(4\varrho_{0}^2 x^2 - a_{0})
\end{equation}
and has a minimum at $x=0$ with $a_{0}>0$, which implies
that the shell will inevitably collapse to a black hole, regardless
of the sign of its initial velocity, and it will shrink to the
origin within finite time.  Choosing initial conditions so that
$\dot{x_0}=0$,  eq. (\ref{eq:re14b}) can be rewritten as
\begin{equation}
  \label{eq:inaddre}
  M = \varrho_{0}\left[2(x_{0}^2+k_-)^{1/2}  - \varrho_{0}\right] -k_-,
\end{equation}
which is analogous to the condition for dust ball collapse
\cite{mr}. A black hole can form only if
\begin{equation}
\label{eq:rhocondition} \varrho_0 > (x_{0}^2+k_-)^{1/2} - x_0
\end{equation}
otherwise $M<0$ and the shell collapses to a naked singularity and a Cauchy horizon
forms. Note
that if $-M < k_- < 0$ then the condition (\ref{eq:rhocondition}) is
always satisfied. It is curious that for sufficiently small initial
shell density that cosmic censorship is violated.

The comoving time for the shell radius to become coincident with the
event horizon is given by (\ref{eq:horcoltime}), and the time $t_c$
for the shell to collapse from $x_{i}=x_0$ to $x_f=x(t_{c}) = 0$ is
always finite for positive $M$ and $\varrho_{0}$. However, the
coordinate time at which an observer outside the black hole observes
the collapse is not finite since
\begin{eqnarray}
  \label{eq:coortime}
  \hat{T} &=& T_{0} + \int_{x_{0}}^{\hat{x}} \frac{\ell
  dx}{(x^2-M)}\nonumber \\ &=& T_{0} - \frac{\ell}{\sqrt{M}} {\rm
  arctanh}\left[\frac{x}{\sqrt{M}}\right]_{x_{0}}^{\hat{x}},
\end{eqnarray}
where $\hat{T}$ is a coordinate time at which a signal emitted from
the edge of the shell arrives at a certain point $\hat{x}$. This
coordinate time is clearly divergent when ${\hat{x}} \rightarrow
\sqrt{M}=x_{H}$, which implies that the collapse to the horizon takes
infinite time, so that observers outside the black hole will not
observe this collapse.

The redshift of light from the edge of the dust shell is
\begin{equation}
  \label{eq:redshift}
  z = \frac{d\hat{T}}{dt} - 1 = \frac{\dot{x}}{(x-x_{H})(x+x_{H})} - 1,
\end{equation}
which obviously diverges at $t=t_{H}$($x=x_{H}$). Thus the
collapsing shell of dust will fade away from observer's sight as
time goes by, as with the collapse of the dust ball \cite{mr}.

Note that if $M=-m<0$ then eq. (\ref{eq:requires}) is not
necessarily satisfied. If $\rho_0>0$ then $1>k>m$, and
the conical deficit angle outside the shell is larger than that
inside the shell. In this case the larger the shell density, the
larger the exterior deficit angle relative to the interior one. The density is bounded by
\begin{equation}
\label{eq:rhocondition2} \varrho_0 < (x_{0}^2+k)^{1/2} - x_0
\end{equation}
which ensures that the exterior deficit angle is always less than
$2\pi$.

We close this subsection by noting that if $k_+>0$ then the
shell can collapse to a naked singularity if the initial conditions
are properly set. The alternative to this scenario is that the shell
either expands indefinitely or collapses to a minimal size and then
expands indefinitely. As noted previously, the specifics depend on
the values of the parameters $a_-$ and $k_{\pm}$.

\subsection{External Flat Space}

Now we turn to shell collapse where Minkowski and/or conical deficit
space describing a point mass is inside and/or outside the shell,
corresponding to the case $a_+=0$.

Consider first the case where $a_-=0$ as well, yielding
$F_{\pm} = k_{\pm}$, representing a flat space with conical deficit
when $0<k_{\pm}<1$ (which vanish when $k_{\pm} = 1$), where
$k_{\pm}>0$ to preserve the sign of the metric. For positive energy
density $k_{+}<k_{-}$, ensuring that the exterior deficit angle is
greater (corresponding to a larger mass) than the interior deficit
angle as noted above. Then the equation of motion (\ref{eq:re14b})
has the solution
\begin{equation}
  \label{eq:solsMS}
  x(t) = x_{0} + \dot{x}_0 t,
\end{equation}
which is analogous to dust ball collapse in Minkowski space
\cite{mr}. The coefficient of $t$ corresponds to the initial
velocity of the disk, which must be negative if collapse is to take
place (if it is positive then the shell expands outward without
resistance). The initial velocity is
\begin{equation}
\dot{x}_0 = \pm \frac{\sqrt{\varrho_{0}^4 -
  2(k_{+}+k_{-})\varrho_{0}^2 +(k_{+}-k_{-})^2}}{2\varrho_{0}},
\end{equation}
which is real provided
\begin{equation}
  \label{eq:kpmrhocond}
    \varrho_{0}>\sqrt{k_{-}}-\sqrt{k_{+}}.
\end{equation}
Note that there is no collapse unless the shell is given
some initial inward velocity, ie ${\dot{x}_0}<0$, in which case the
shell will collapse to a naked singularity with its associated Cauchy horizon\footnote{Note that even if the energy
density of the shell is negative, collapse can still take place
if ${\dot{x}_0}<0$. In this case the exterior deficit angle is
smaller than the interior one.}.

Next we consider the more general case of an interior
spacetime with cosmological constant. Without loss of generality we
can set $a_-=\pm 1$.  The effective potential is still a quartic
given by eq. (\ref{eq:effp}) and the solution is given by
(\ref{eq:sol14b}), both with $a_+=0$ and $|a_-|=1$. Positivity of
the initial density $\varrho_{0}>0$ implies
\begin{equation}
\label{eq:inis2} \sqrt{a_- x_{0}^2+k_{-}}-\sqrt{k_{+}}
>0 \longrightarrow a_- x_{0}^2 > k_{+} - k_{-}
\end{equation}
and so either $k_+>0$ must be sufficiently small if the interior is
AdS or else $k_->0$ must be sufficiently large if the interior is
dS. In either case the shell will either expand indefinitely,
undergo a bounce after which it expands indefinitely, or else
collapse to a naked singularity.

\subsection{External dS Space}

Finally, we assume that the exterior metric function is
that for a dS spacetime, ie. $a_+=-1$, yielding $F_{+}=(\mu -
{\mathcal R}^2/\ell^2)$. The cosmological horizon is located at
${\mathcal R}_{h}=\sqrt{k_+}\ell$, where $k_+ = \mu >0$ in order to
have the metric signature correct. The exterior metric is that of dS
spacetime with a conical deficit unless $\mu=1$, in which case it is
pure dS spacetime.

The general solution is again given by eq.
(\ref{eq:sol14b}), where
\begin{equation}
  \label{eq:sol14b-dS}
 x(t)=-\frac{\gamma_{-}}{|1 + a_{-}|}
 {\rm JacobiSN}\left[\frac{\gamma_{+}}{2\varrho_{0}}(t-t_{0}),
 {\frac{\gamma_{-}}{\gamma_{+}}}\right]
\end{equation}
with
\begin{equation}
  \label{eq:sol14bconstAdS}
a_4=(1+a_{-})^2 \qquad a_2=-(1-a_{-})\varrho^2_{0}+(a_{-}+1)(\mu -
k_{-}) \qquad a_0=\left(\varrho^2_{0}-(\mu +k_{-})\right)^2 - 4 \mu
k_{-}.
\end{equation}
and again $\gamma_{\pm}$ is given by eq. (\ref{eq:sol14bconsts}).

Positivity of energy now imposes the requirement
\begin{equation}
\label{eq:inis3}  (a_{-}+1) x_{0}^2 > \mu - k_{-}
\end{equation}
which implies that $\mu$ must be sufficiently small relative to the
other parameters.  If the interior is AdS, then $a_->0$ and the
initial shell size $x_0$ must be sufficiently large relative to the
sum of the masses (if there is a black hole in the interior, with
$k_-=-M<0$) or their difference (if there is a point mass in the
interior, with $0<k_-<1$). The same analysis holds true if the
interior is a dS space with smaller cosmological constant (ie. $-1 <
a_- < 0$), though in this case $k_->0$.  If $a_- <-1$ then $k_-$
must be sufficiently large for the shell to have any allowed motion.

The possible trajectories of the shell have been covered at
the beginning of this section.  If the shell initially contracts, it
will either collapse to a naked singularity or else it will bounce
at some finite radius
\begin{equation}
  \label{eq:dsexpand}
  x_{min}
=\frac{\sqrt{-(1-a_{-})\varrho^2_{0}+(1+a_{-})(\mu-k_{-})+2\varrho_{0}\sqrt{-a_{-}(\varrho^2_0
-\mu - k_-) + k_{-} + a^2_{-}\mu}}}{|1+a_{-}|}
\end{equation}
and then expand to indefinitely large size. If the shell initially
expands, it will either expand indefinitely or it will bounce at
\begin{equation}
  \label{eq:dsexpand2}
  x_{max}
=\frac{\sqrt{-(1-a_{-})\varrho^2_{0}+(1+a_{-})(\mu-k_{-})-2\varrho_{0}\sqrt{-a_{-}(\varrho^2_0
-\mu - k_-) + k_{-} + a^2_{-}\mu}}}{|1+a_{-}|}
\end{equation}
and then collapse to a naked singularity. The remaining
alternative is that of a shell in unstable equilibrium, which occurs
if $ -a_{-}(\varrho^2_0 -\mu - k_-) + k_{-} + a^2_{-}\mu$ and if the
initial conditions are properly set. Collapse to a black hole is
never possible.

\section{Collapse of a Shell with Pressure}\label{sec:pressure}

We now consider the collapse of a shell with pressure, whose
equation of state we take to be that of a polytrope
\begin{equation}
\label{eq:fluid} p=q\rho\left(\frac{\rho}{\rho_0}\right)^{1/n}
\end{equation}
where $q$ is a constant, representing diverse choices for the matter
content of the shell \cite{cf}. For example, $n=0$, $n=1$, and $n=2$
repectively represent constant energy density, nonrelativistic
degenerate fermions, and nonrelativistic matter or  radiation
pressure. Moreover, the equation of state for perfect fluids is
achieved by setting $n\rightarrow \infty$.

The matching conditions (\ref{eq:6}) and (\ref{eq:7}) of the shell
imply that
\begin{equation}
\label{eq:fluidmatch} \frac{d}{d{\mathcal R}}\left(\rho {\mathcal
R}\right) = -p = -q\rho^{1+1/n}
\end{equation}
which yields
\begin{equation}
\label{eq:fluidmatchsol} \rho (\mathcal{R}) =
\rho_{0}\left(-q+K\left(\frac{\mathcal{R}}{\ell}\right)^{1/n}\right)^{-{n}}
\end{equation}
where $K$ is a  constant of integration. For the perfect fluid, we
obtain
\begin{equation}
\label{eq:fluidperfectmatchsol} \rho (\mathcal{R}) =
\rho_{0}\left(\frac{\ell}{\mathcal{R}}\right)^{1+q}
\end{equation}
From these we respectively obtain the equations
\begin{equation}
  \label{eq:flushell}
  \sqrt{\dot{x}^2 + a_{+}x^2 + k_{+}} -
  \sqrt{\dot{x}^2 + a_{-}x^2 + k_{-}} + \frac{\varrho_{0} x}{(-{q}+(1+q)x^{1/n})^{n}} = 0,
\end{equation}
for finite $n$, and
\begin{equation}
  \label{eq:perflushell}
  \sqrt{\dot{x}^2 + a_{+}x^2 + k_{+}} -
  \sqrt{\dot{x}^2 + a_{-}x^2 + k_{-}} + \frac{\varrho_{0}}{x^{q}} = 0,
\end{equation}
for the perfect fluid, where again $x \equiv {\mathcal R}/\ell$,
$\varrho_{0}=\rho_{0}{\ell}$, and $K=(1+q)$ so that $\rho=\rho_0$
when $x=1$.

For all $n$, the equation of motion can be written as
\begin{equation}
\dot{x}^2 + V_{\rm eff}(x) = 0 \label{eq:eqnpres}
\end{equation}
where
\begin{equation}
 V_{\rm eff}(x) = -\frac{x^2}{4}\left(\rho^2-2\left(a_{+}-a_{-}\right)+\frac{\left(a_{+}-a_{-}\right)^2}{\rho^2}\right) +\frac{k_{+}-k_{-}}{2}
 - \frac{\left(a_{+}-a_{-}\right)\left(k_{+}-k_{-}\right)}{\rho^2} +\frac{\left(k_{+}-k_{-}\right)^2}{4\rho^2 x^2}
\end{equation}
is the effective potential. In general, it depends on many
parameters, ($\varrho_0$, $q$, $a_{\pm}$, $k_{\pm}$) and so is
somewhat unwieldy to analyze in full generality.  Furthermore it is
hard to obtain an exact solution of eq. (\ref{eq:flushell}) since
the equation of motion is highly nonlinear.

However the structure of the effective potential allows us to
discern some basic features.  First, for finite $n$ and nonzero $q$,
the shell will not collapse to a point, but rather to a ring of size
$x=x_q\equiv q^n/(1+q)^{n}$ in finite proper time. Physically we can
think of the shell as developing an increasingly large internal
pressure that diverges for some finite value of the shell radius.
For the perfect fluid the shell can collapse to zero size in finite
proper time. In either case the stress energy tensor of the shell
diverges.  However there is no backreaction since the field
equations force spacetime to have constant curvature in regions
where the stress-energy vanishes, ie. everywhere outside of the
shell.

However the intrinsic Ricci scalar can be written in the form
\begin{equation}
R_{\mu}^{\mu}[\Sigma] = \frac{1}{x^2}\left[x\dot{x}^2 -
\ddot{x}\right] = \frac{1}{2x^2}\left[\frac{d}{dx}V_{\rm eff}(x) -
2xV_{\rm eff}(x)\right]
\end{equation}
by using eq. (\ref{eq:eqnpres}). Since $V(x_q)\rightarrow -\infty$
and $dV(x_q)/dx \rightarrow \infty$, the intrinsic Ricci scalar of
the shell diverges at $x=x_q$ as the shell approaches its minimal
radius. Physically the internal pressure forbids the shell
to be compressed without limit. As it shrinks in size, the pressure
grows, eventually diverging (along with the density) at some finite
shell radius.

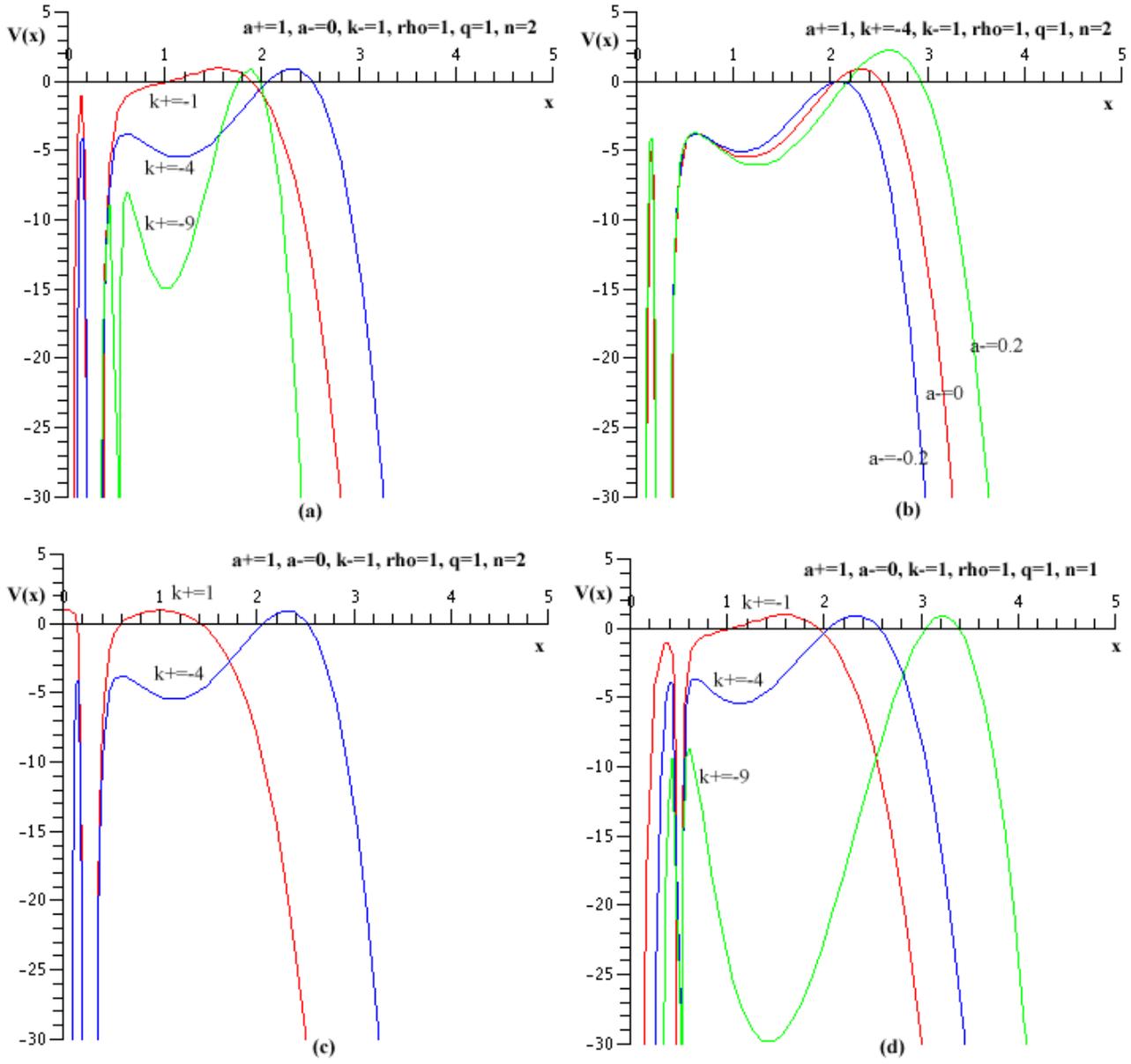
\fig{17cm}{16cm}{pressureffpts}{\small Some plots of the effective
potential of the shell with pressure. The parameters have been
chosen so that the pressure and density diverge at $x=1/2$; the
effective potential is physically meaningful only for $x>1/2$.
}{fig:pressurepts}

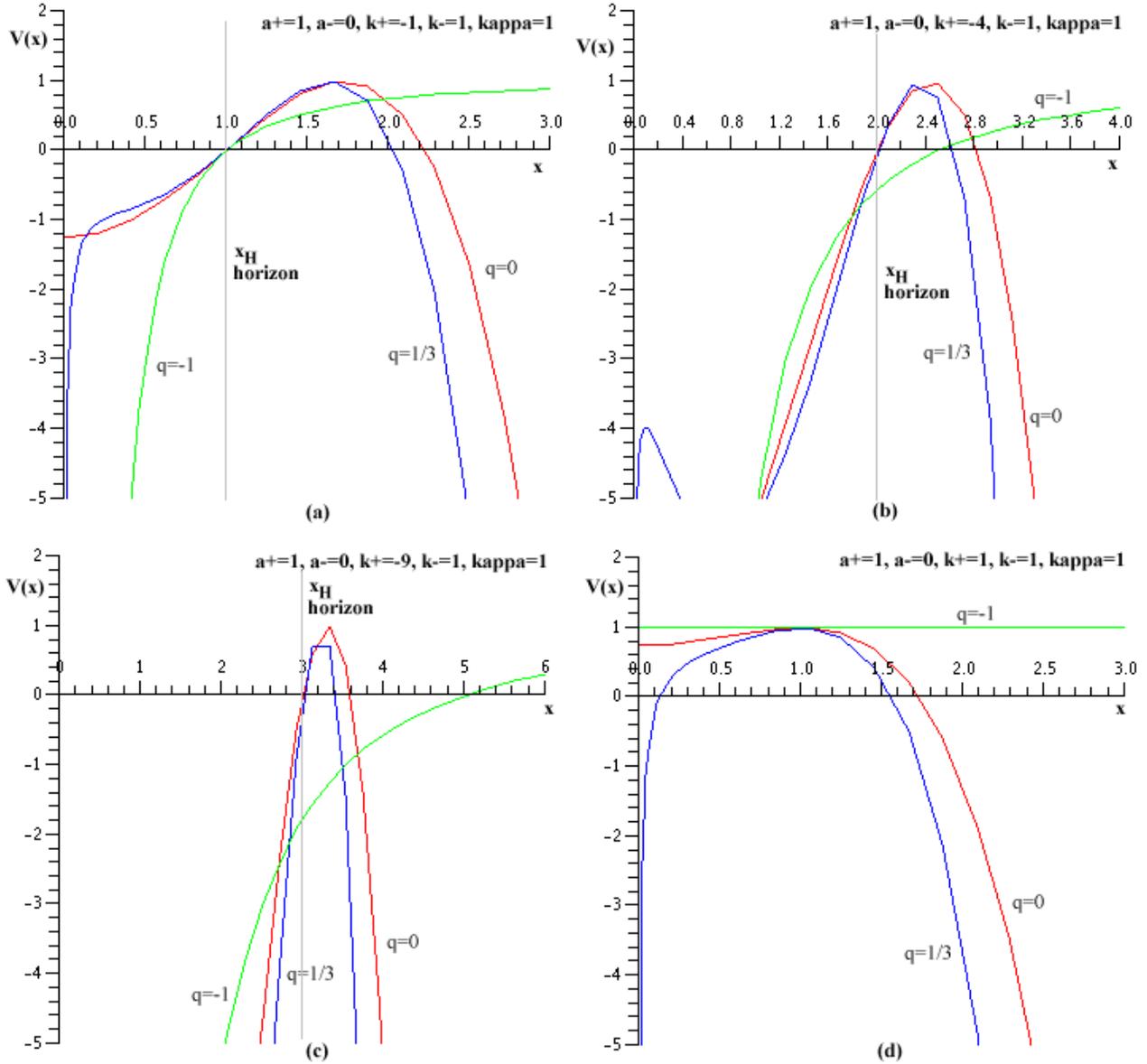
\fig{17cm}{16cm}{effptpfluid}{\small Some plots of the effective
potential of the shell of perfect fluids, for example, $q=0$ (dust),
$q=1/3$ (radiating matter), and $q=-1$ (cosmological
constant).}{fig:effptperfectfluid}

The shell can exhibit  several kinds of behavior, depending upon the
values of the parameters, $a_{\pm}$, $k_{\pm}$, $q$, $n$, and
$\varrho_0$. We illustrate here the generic possibilities for the
effective potential for specific  values of these constants in Fig.
\ref{fig:pressurepts} (for perfect fluids in Fig.
\ref{fig:effptperfectfluid}).  Generically the effective potential
has 2 local maxima and one local minimum, and it diverges to minus
infinity for large $x$ and for some finite $x$.  It is possible for
the rightward local maximum to occur for positive values of $V_{\rm
eff}$, in which case the shall expands to infinity, possibly after a
bounce if it is given inward initial velocity.  The event horizon
for the black hole is always to the left of this maximum (and occurs
where $V_{\rm eff}<0$) so for initial values of the shell radius
between the event horizon and the smaller root of $V_{\rm eff}$ the
shell will always collapse into a black hole, again possibly with a
bounce if given outward initial velocity. If the rightward local
maximum of $V_{\rm eff}$ occurs at $V_{\rm eff}<0$ then the shell
will either expand outward to infinity or collapse to a black hole
depending on whether the initial velocity is outward or inward.
Variation of the parameters can cause the local minimum to
disappear, leaving a single maximum for the effective potential, in
which case the same qualitative behaviour of the shell takes place
as previously described.  A numerical search indicates that there
are no values of the parameters for which both local maxima occur
for $V_{\rm eff}>0$, and so the shell can never undergo bouncing
oscillations between maximum and minimum values.

However it is possible for the shell to collapse without forming an
event horizon if $k_{+}>0$. In this case the pressure and density
diverge in finite proper time at some finite value of the shell
radius (or zero value in the case of a perfect fluid) as discussed
above. The stress-energy tensor and the intrinsic Ricci scalar of
the shell both diverge, and it is not possible to evolve the shell
beyond this point.  In this sense we have a mild violation of cosmic
censorship; although the curvature is finite everywhere outside the
shell, it diverges on the shell, as with the case of pressureless
dust.  The pressure, rather than preventing a singularity, instead
moves it out to finite shell radius.

\section{Collapse of the Generalized Chaplygin Shell}
\label{sec:GCG}

We now turn to consideration of the gravitational collapse of a
generalized Chaplygin gas (GCG) shell with an equation of state,
$p=-\tilde{A}/\rho^{\alpha}=-A\rho_0\left(\rho_0/\rho\right)^\alpha$.
We find that the density is
\begin{equation}
  \label{eq:gcrho}
\rho=\frac{[A\rho^{\alpha+1}_0{\mathcal
R}^{\alpha+1}+C]^{\frac{1}{\alpha+1}}}{{\mathcal R}}
\end{equation}
where $C$ is a constant of integration. Combining eqs. (\ref{eq:6})
and (\ref{eq:7}) leads to a simple equation,
\begin{equation}
  \label{eq:gcgeq}
\sqrt{\left(\frac{d{\mathcal R}}{d\tau}\right)^2 + F_{+}} -
\sqrt{\left(\frac{d{\mathcal R}}{d\tau}\right)^2 + F_{-}} +
[A\rho^{\alpha+1}_0{\mathcal R}^{\alpha+1}+C]^{\frac{1}{\alpha+1}} =
0,
\end{equation}
where $C$ has been chosen so that $\rho=\rho_0$ when $x\equiv{\mathcal R}/\ell = 1$.

The setting is like that of a polytrope, but with $q<0$ and
$n={1}/(1+\alpha)$. If we set $A=0$ the situation reduces to
that of the collapse of pressureless dust shell investigated in the
previous section.  If $A> 0$  eq. (\ref{eq:gcgeq}) describes a GCG
with a negative pressure. Provided $\alpha>-1$, the density will
diverge at the origin for $C>0$. If $C<0$ it will converge to some
finite value at some nonzero value of the shell radius ${\mathcal
R}$, with the pressure diverging at that same radius. If $C=0$ the
density and pressure are constant for all values of ${\mathcal R}$.

We proceed as before by redefining parameters so that  $x \equiv
{\mathcal R}/\ell$, $\varrho_0 \equiv \rho_0\ell$,  $\tau \equiv
t/\ell$, and set $C= 1-A \varrho_0^{(\alpha+1)}$, so that
$\rho=\rho_0$ when $x=1$. Then eq. (\ref{eq:gcgeq}) becomes
\begin{equation}
  \label{eq:eq2}
  \sqrt{\dot{x}^2 + a_{+}x^2 + k_{+}} -
  \sqrt{\dot{x}^2 + a_{-}x^2 + k_{-}} + \varrho_0 [A x^{\alpha+1} +
  1-A]^{\frac{1}{\alpha+1}} = 0.
\end{equation}
The equation of motion can be rewritten as
\begin{equation}
\dot{x}(t)^2 + V_{\rm eff}(x)=0,
\end{equation}
where the effective potential is
\begin{equation}
V_{\rm eff} (x) = \frac{1}{4{\mathcal C}^2} \left[ \left({\mathcal C}^2-((a_{+}+a_{-})x^2+k_{+}+k_{-})\right)^2 - 4(a_{+}x^2+k_{+})(a_{-}x^2+k_{-})\right]
\end{equation}
with ${\mathcal C}=\varrho_0(A x^{\alpha+1}+1-A)^{1/(\alpha+1)}$,
whose shape depends upon several parameters, $\alpha$, $A$,
$a_{\pm}$, and $k_{\pm}$. Note that the effective potential
will not diverge for any finite value of $x$ if $A<1$.
Otherwise, for all $\alpha>-1$ it will diverge quadratically for
large $x$ to either positive or negative infinity, depending on the
values of the parameters.

For example, $\alpha=1$ describes a Chaplygin gas shell, whose effective potential is
\begin{equation}
V_{\rm eff}(x) = -
\frac{1}{4\varrho_0(\omega^2x^2+c)}(a_4x^4-2a_2x^2+a_0),
\end{equation}
where for convenience we have defined
$\omega^2=A\varrho_0^2$ and $c=(1-A)\varrho_0^2$ so that
\begin{eqnarray}
&&a_4=(\omega^2 - (a_{+}+a_{-}))^2 -4a_{+}a_{-},\nonumber \\
&&a_2=(k_{+}+k_{-}-c)\omega^2 + (a_{+}+a_{-})c-(k_{+}-k_{-})(a_{+}-a_{-}),\nonumber\\
&&a_0=(c-(k_{+}+k_{-}))^2 -4k_{+}k_{-}.
\end{eqnarray}
The effective potential for arbitrary $c$ is plotted in Fig.
\ref{fig:gcgeffptC} for an exterior AdS black hole metric
($k_{+}<0$) and an exterior AdS point mass metric ($k_{+}>0$)
outside the shell.

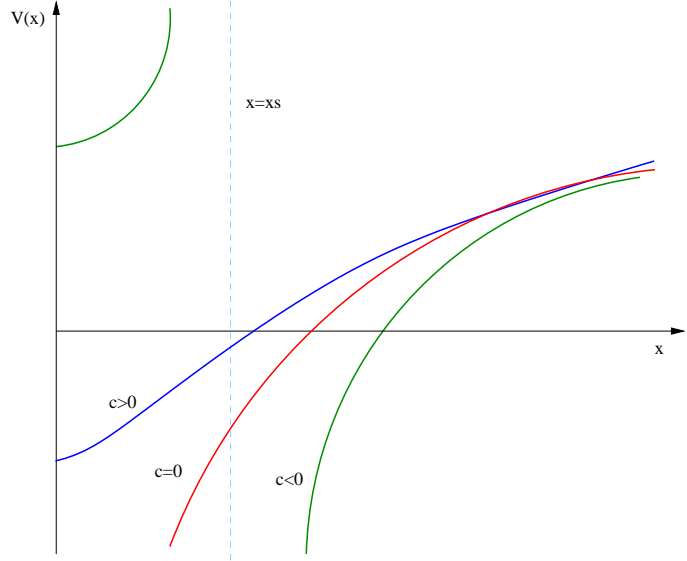
\fig{9cm}{7.5cm}{gcgeffptC}{\small Plot of the effective potential
of the GCG shell with arbitrary $c$ for $\alpha=1$, where $x_s
=\varrho_0\sqrt{1-1/A} = \sqrt{-c}/\omega$ only for
$c<0$.}{fig:gcgeffptC}

 A glance at Fig. \ref{fig:gcgeffptC} indicates that the
GCG shell will collapse to an AdS black hole within a finite time
for $c\ge 0$ while this is always not the case if $c<0$. If
$x_{H}>x_{s}$, the shell will form a black hole while if
$x_{H}<x_{s}$, it will collapse to a finite size of $x=x_{s}$, even
if the exterior metric is a BTZ black hole. The endstate is
a rather unusual state in which the density is finite but the
pressure diverges, again yielding a Cauchy horizon and a violation
of cosmic censorship.

Consider the special case $C=0$. This leads to the simpler form
\begin{equation}
  \label{eq:eq3}
  \sqrt{\dot{x}^2 + a_{+}x^2 +k_{+}} -
  \sqrt{\dot{x}^2 + a_{-}x^2 +k_{-}} + \omega x = 0,
\end{equation}
or alternatively
\begin{equation}
  \label{eq:altereq}
  \dot{x}^2 + V_{\rm eff}(x) = 0,
\end{equation}
where $V_{\rm eff}(x)$ is an effective potential with
\begin{equation}
\label{eq:effptgcg}
V_{\rm eff}(x) = - \frac{1}{4\omega^2 x^2}\left[a_{4}x^4 - 2a_{2}x^2 + a_{0}\right],
\end{equation}
where
\begin{equation}
\label{eq:acoeffgcg}
a_{4} = (\omega^2-a_{+}-a_{-})^2-4a_{+}a_{-} \qquad
a_{2} = (k_{+}+k_{-})\omega^2 - (k_{+}-k_{-})(a_{+}-a_{-})\qquad
a_{0} = (k_{+}-k_{-})^2.
\end{equation}

For this case the density and pressure and always
constant. The specific shape of the potential will depend upon the
values of $a_4$, $a_2$, and $a_0$ associated with $a_{\pm}$,
$k_{\pm}$, and $\omega$. For example, if $a_{0}=0$ (ie.
$k_{+}=k_{-}$), it is simply described by the usual quadratic
function. The shape of the effective potential for each case is
plotted in Fig. \ref{fig:effectivept}.

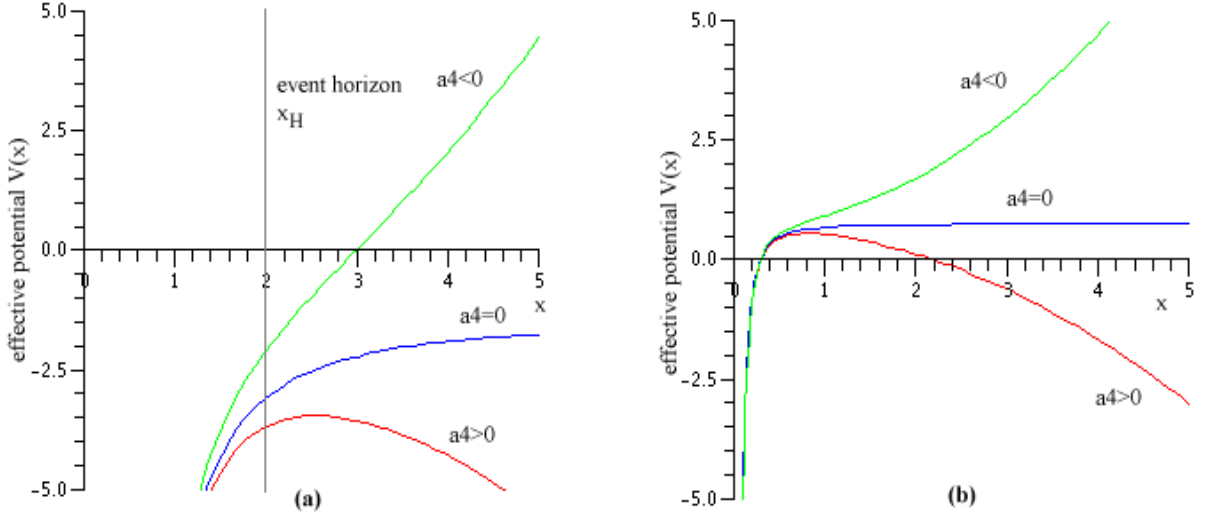
\fig{16cm}{7cm}{effpta4C0}{\small Plots of the effective potential
of the GCG shell for each $a_4$, setting $\alpha=1$, $k_{-}=1$, and
$k_{+}=-4$ (LHS) and $k_{+}=0.5$ (RHS).}{fig:effectivept}

For $a_0>0$, the behavior of $x\rightarrow 0$ is $V_{\rm
eff}(x\rightarrow0)=-\infty$, regardless of the value of $a_4$. For
$a_4>0$, since the potential has two roots at
$x=\sqrt{a_2\pm\sqrt{a_2^2-a_4a_0}}/\sqrt{a_4}$ and $V_{\rm
eff}(x\rightarrow \infty) = -\infty$, the curve of the potential is
concave down. In this case the shell will expand  to infinity for
sufficiently large initial radius (with a bounce if the initial
velocity is negative) or collapse to zero size (leaving behind a
black hole or a naked singularity) for sufficiently large initial
radius (with a bounce if the initial velocity is positive). For
$a_4<0$  there is only one root and $V_{\rm eff}(x\rightarrow\infty)
= \infty$. In this case the shell always collapses to zero size,
possibly preceded by a bounce if it is initially expanding outward.
For $a_4=0$ a third possibility exists in which the shell can expand
indefinitely but will continually decelerate. These
scenarios are depicted on the left-hand-side of figure
\ref{fig:effectivept}.

For $k_{+}<0$ (an exterior AdS black hole metric) and $a_4\ge 0$
(blue and red lines in Fig. \ref{fig:effectivept} (a)), the shell
can either collapse or expand, depending upon the initial velocity,
$\dot{x}$. However for $a_4<0$, the shell will either collapse to a
black hole or initially expand, bounce and then collapse to a black
hole. For $k_{+}>0$ (an AdS point mass  metric) and $a_4 \le 0$
(green and blue lines in Fig. \ref{fig:effectivept} (b)), the shell
will collapse to a point at $x=0$ while if $a_4>0$, then the shell
will either collapse to a point or expand indefinitely or collapse
to a certain size and expand indefinitely again.

There are three classes of solutions,  depending upon the sign of
$a_4$: $a_4>0$, $a_4=0$, and $a_4<0$. The general solutions of eq.
(\ref{eq:eq3}) are
\begin{eqnarray}
x(t)^{a_{4}>0}&=&\frac{1}{\sqrt{a_4}}\left[a_2\left(1-{\rm cosh}\left[\frac{\sqrt{a_4}}{\omega}(t-t_0)\right]\right)+\sqrt{a_4a_0}{\rm sinh}\left[\frac{\sqrt{a_4}}{\omega}(t-t_0)\right] \right]^{\frac{1}{2}}\label{eq:gensola4ge0}\\
x(t)^{a_{4}<0}&=&\frac{1}{\sqrt{\bar{a}_4}}\left[\sqrt{a_2^2+a_0\bar{a}_4} \sin\left[{\rm arcsin}\left(\frac{a_2}{\sqrt{a_2^2+a_0\bar{a}_4}}\right)-\frac{\sqrt{\bar{a_4}}}{\omega}(t-t_0) \right]-a_2 \right]^{\frac{1}{2}}\label{eq:gensola4le0}\\
x(t)^{a_4=0}&=& \sqrt{\frac{a_{0}\omega^2 - a_{2}^2 \left(t-t_{0}+\frac{\sqrt{a_0}\omega}{a_2}\right)^2}{2 a_{2}\omega^2}}\label{eq:gensola40}
\end{eqnarray}
where $t_0$ is the collapse time to $x=0$, respectively given by
\begin{eqnarray}
&&t_0^{a_4>0} = \frac{\omega}{\sqrt{a_{0}}} {\rm arccosh}\left[\frac{a_2(a_4x_0^2-a_2)+ a_4\sqrt{a_0(a_4x_0^4-2a_2x_0^2+a_0)}}{a_0a_4-a_2^2} \right] \label{eq:colta4ge0}\\
&&t_0^{a_4<0} = \frac{\omega}{\sqrt{\bar{a}_4}} \left[{\rm arcsin}\left[\frac{a_2+\bar{a}_4 x_{0}^2}{\sqrt{a_2^2+a_0\bar{a}_4}}\right]-{\rm arcsin}\left[\frac{a_2}{\sqrt{a_2^2+a_0\bar{a}_4}}\right]\right]\label{eq:colta4le0}\\
&&t_0^{a_4=0} = \frac{\omega}{a_2} (\sqrt{a_0}-\sqrt{a_0-2a_2x_0^2}),\label{eq:colta40}
\end{eqnarray}
where $\bar{a}_4 = - a_4 > 0$.

In order to have a positive and real collapse time in eq. (\ref{eq:colta4ge0}), we should impose that the argument of cosine hyperbolic function must be greater than $1$, which leads to a condition,
\begin{equation}
a_0a_4-a_2^2 > 0,
\end{equation}
while eqs. (\ref{eq:colta4le0}) and (\ref{eq:colta40}) is always valid regardless of parameters.

The intrinsic scalar curvature of the shell can be evaluated by an expansion in terms of $t-t_0$ for each case,
\begin{equation}
R_{\mu}^{\mu}[\Sigma] \sim -\frac{3}{8(t-t_0)^2} \cdots  \label{eq:riccia4g0}\\
\end{equation}
where we see that it is generically singular at the endpoint of collapse.

An interesting subcase is obtained by  setting  $k_{+}=k_{-}=k$,
which yields $a_4=(\omega^2-a_{+}-a_{-})^2 - 4a_{+}a_{-}$,
$a_2=2k\omega^2$, and $a_{0}=0$. The effective potential is now a
quadratic function in the form of
\begin{equation}
V_{\rm eff}(x) = - \frac{1}{4\omega^2}(a_4x^2-2a_2),
\end{equation}
and possible scenarios of collapse, depending upon $a_4$ and
$a_2$, are shown in Fig. \ref{fig:kpkmequal}.
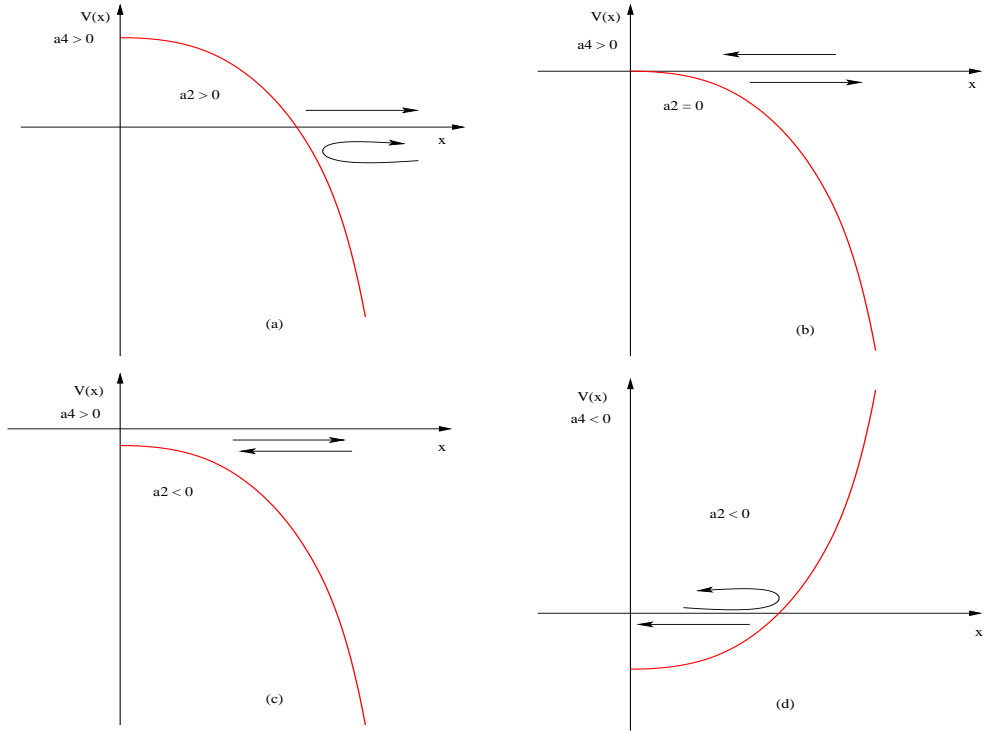
\fig{13cm}{9.7cm}{kpkmeq}{\small Possible collapse scenarios of
$k_{+}=k_{-}$ case.}{fig:kpkmequal} For $a_4>0$, there are three
choices of collapse scenarios depending on the sign of $a_2$. For
an AdS space outside the shell ($a_{+}=1$), then
$a_4=(\omega^2-1-a_{-})^2-4a_{-} >0$. When $a_2>0$, then $k>0$,
which implies that we have the metric of an AdS point mass outside
the shell. The interior space is either (i) an AdS point mass if
$a_{-}>0$, (ii)  a dS space if  $a_{-}<0$, or (iii) a point mass
in a flat space if $a_{-}=0$. Then the shell will either bounce at
$x=\sqrt{2a_2/a_4}$ and expand again if it initially contracts or
expand indefinitely, as shown in Fig. \ref{fig:kpkmequal} (a). If
$a_2=0$, then $k=0$. Thus $a_{-}>0$  to preserve the metric
signature, which describes AdS vacuum in and outside of the shell.
Then the shell will either expand indefinitely or contract
to zero size, at which point its intrinsic Ricci scalar diverges,
forming a naked singularity (see \ref{fig:kpkmequal} (b)).

Alternatively the situation $a_2<0$ describes two AdS black holes
with $k=-M$ since $a_{-}>0$ for preserving the metric signature.
Note that these holes will have differing masses since $a_{-}>
a_{+}$. In this case, the shell will either collapse to a black hole
or expand, depending upon its initial motion. (Fig.
\ref{fig:kpkmequal} (c)) Finally, for $a_4<0$, it is found that
$a_{+}a_{-}>0$, ie.  both spaces should be AdS spaces. Since
$a_2=2k\omega^2 <0$, this also describes AdS black holes of
differing mass, even if the shape of the effective potential is
different from above case.

Note that the endstate of collapse for all black hole
scenarios is a singularity cloaked by an event horizon. Since
$k_{+}=k_{-}$,  we obtain from eqs. (\ref{eq:gensola4ge0}) and
(\ref{eq:gensola4le0})
\begin{eqnarray}
&&R_{\mu}^{\mu}[\Sigma]^{a_4>0} = - \frac{a_4}{8\omega^2}\left[ 1- \frac{2}{{\rm cosh}\left[\frac{\sqrt{a_4}}{\omega}(t-t_0)\right]-1}\right],\\
&&R_{\mu}^{\mu}[\Sigma]^{a_4<0} = - \frac{a_4}{8\omega^2}\left[ 1-
\frac{2}{{\cos}\left[\frac{\sqrt{a_4}}{\omega}(t-t_0)\right]-1}\right],
\end{eqnarray}
for the intrinsic curvature scalars in each case, which are both
singular as $t\rightarrow t_0$. Consequently (Figs.
\ref{fig:kpkmequal} (c) and (d)) once the shell hits zero size a
singularity is hit and then we lose predicability.

Now let us consider a point mass in flat space outside the shell.
Then we have $a_{+}=0$ and $k>0$, which implies $a_4>0$ and $a_2>0$,
regardless of the sign of $a_{-}$. Then the shell will ultimately
expand in this case.

If there is a dS space outside the shell (ie. $a_{+}<0$ and $k>0$),
we also have $a_2>0$ while $a_4$ can have both signs, depending upon
the choice of $a_{-}$. If $a_{-}\ge 0$ (a point mass in AdS or flat
space), $a_4>0$ and this describes an expanding shell as shown in
Fig. \ref{fig:kpkmequal} (a).  However, if $a_{-}<0$, it describes
dS spaces in-and-outside the shell, which leads to an indefinitely
expanding shell. (Fig. \ref{fig:kpkmequal} (a))

\section{Discussion}
\label{sec:discussions}

Perhaps the most intriguing result of this paper is that shell
collapse in $(2+1)$-dimensional gravity can violate --albeit
somewhat mildly -- cosmic censorship for a broad range of initial
conditions, whether we have pressureless dust shells, shells with
pressure, or GCG shells. The situation is markedly different from
that of a scalar field \cite{scalarcollapse}, in which either a
black hole is formed or the scalar field oscillates indefinitely
without collapse. Here, as the shell collapses its density (and
pressure, if any) diverge in finite proper time.  Although the
exterior spacetime develops no curvature singularities (since
spacetime in (2+1) dimensions has constant curvature outside of all
matter sources), the stress-energy of the shell diverges in finite
proper time and so the Einstein equations (and the second junction
condition in eq. (\ref{eq:junctions})) break down. Consequently the
equation of motion describing the time-evolution of the shell is no
longer valid. This yields a mild violation of cosmic censorship in
that, strictly speaking, there is no definite manner in which to
continue the spacetime beyond this event and its future light
cone\footnote{For a discussion of possible violations of cosmic
censorship in fluid collapse in higher dimensions, see
\cite{hdfluid}.}. A Cauchy horizon forms if this endstate is not
cloaked by an event horizon.

We find that dust shells can collapse to zero size in an AdS
background, displaying similar behavior to that of a pressureless
disk of dust \cite{mr}. The endstate of collapse in both cases is
one in which the shell/disk has finite velocity when it achieves
zero size, with a diverging intrinsic Ricci scalar.  Unlike the
situation in higher dimensions, this state need not be cloaked by an
event horizon, in which case a Cauchy horizon is present.  However
one might take the viewpoint that it is natural to consider matching
this spacetime to one in which the shell bounces repeatedly from
zero size to a maximal value and back again; the exterior space will
always be that corresponding to an AdS point mass, with the interior
space being one of several possibilities as outlined in the
discussion in section 3. The collapse time is always finite. If a
black hole is formed, the edge of the shell and the event horizon
coincide in finite proper time, $t_{h}$.

For shells with pressure, the situation is more intriguing. In this
case the endstate of shell collapse has a finite radius, since the
material of the shell is no longer infinitely compressible. The
intrinsic Ricci scalar becomes singular in finite proper time and
the effective potential diverges. In other words, the shell
collapses to a singular ring with a finite size within finite proper
time. If the external geometry is initially that of a black hole,
that singular ring will be screened by an event horizon. However
this need not be the case, and a naked singular ring with a finite
size can be formed if the exterior metric is that of a point mass.
The role of pressure is that of shifting the singular point at $x=0$
to some finite radial position, sustaining the shell with a finite
size.

Despite the qualitatively different physics of the GCG shell, we
find that it can exhibit similar behaviour to the other two cases.
It also presents a variety of scenarios (such as a collapse to a
black hole or an indefinitely expanding shell) which depend upon the
initial velocity and the shape of the effective potential. However
there are some qualitative differences. It is possible for a GCG
shell to collapse to a shell of finite radius in which the density
is finite but the pressure diverges.  There is still a curvature
singularity on the surface of the shell at the radius $x=x_s$. Even
if the initial external geometry is a black hole, the singular ring
need not be cloaked by an event horizon, and scenarios similar to
the previous cases ensue.

While collapsing shells in $(2+1)$ dimensions are not easily
translatable into realistic scenarios in $(3+1)$ dimensions, our
study is of more than passing interest.  From a general relativistic
viewpoint shell collapse highlights the importance of understanding
what limits there may be to cosmic censorship. Indeed, since there
are two possible endstates for collapse (either a black hole or not)
then there should be some kind of critical phenomenon associated
with this scenario as with the scalar field \cite{scalarcollapse}.
In this context it would be interesting to extend our results to
rotating black holes, where it has been shown that cosmic censorship
holds in dust collapse for the addition of a small amount of angular
momentum \cite{Olea}. From a string-theoretic viewpoint it would be
interesting to understand the implications of this work for the
AdS/CFT correspondence.

\vspace{1cm}

\textbf{Acknowledgments}

This work was supported in part by the Natural Sciences \&
Engineering  Research Council of Canada.  JJO was supported by the
Korea Research Foundation Grant funded by the Korean
Government (MOEHRD: KRF-2005-214-C00148). RBM would like to thank V.
Hubney and JJO would like to thank S. P. Kim, H.-U. Yee, M. I. Park, G. Kang, S.-J. Sin, and K. Choi for fruitful
discussions.

\vspace{1cm}
\newpage

\appendix
\begin{center}
{\bf \Large APPENDIX}
\end{center}
\section{Penrose Diagrams}\label{sec:penrose}

We  consider here the construction of Penrose diagrams for
the various collapse scenarios. With no loss of generality, the
positivity of energy density gives rise to
\begin{equation}
\beta_{-}-\beta_{+} > 0, \label{eq:positivity}
\end{equation}
where $|\beta_{\pm}| = \sqrt{\dot{x}^2+F_{\pm}}$ can each be either
positive or negative and $F_{\pm} = a_{\pm} x^2 + k_{\pm}$.
A wide variety of collapse scenarios are possible under the
constraint of eq. (\ref{eq:positivity}). For example, if $k_{+}>0$
and $k_{-}<0$ (a black hole inside and a point mass outside the
shell), taking  $\beta_{+}>0$ for an expanding shell restricts  $\beta_{-}>0$
but otherwise provides no further constraints.

Cases with an AdS-exterior and dS-interior in the context
of inflationary models have been treated before \cite{MyHub}.  We
shall therefore not consider this case, and concentrate only on a
few of the remaining scenarios. While we have been primarily
interested in shell collapse in this paper, we shall consider
scenarios where the shell can expand as well.

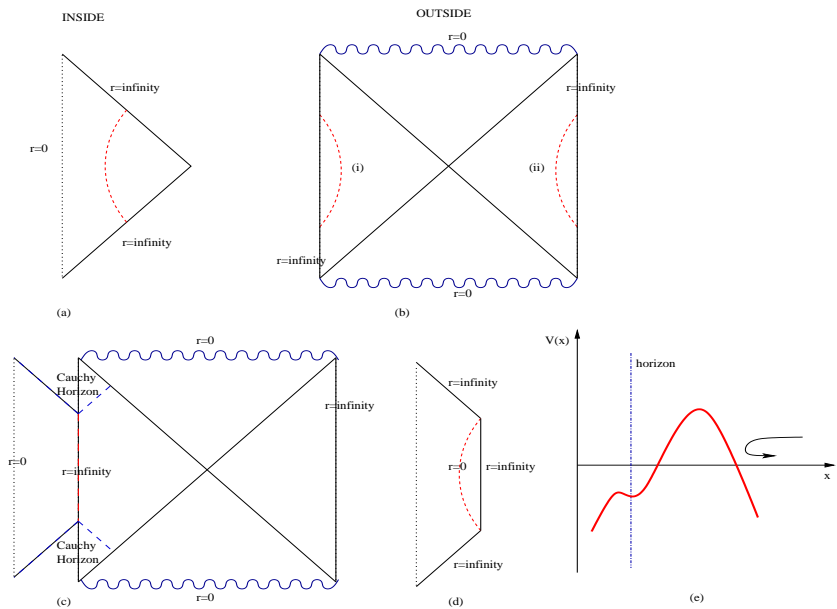
\fig{11cm}{8cm}{penrose1}{\small Penrose diagrams with flat space
inside and a black hole outside the expanding shell. Fig. (a)
represents a flat geometry with a point mass and shell trajectory is
shown in red line for the effective potential (Fig. (e)). The
external geometry has two possibilities of sign of $\beta_{+}$ ((i)
and (ii)). Combined figures of both cases are shown in Figs. (c)
(for (i)) and (d) (for (ii)). Note that for case (c) there is a
Cauchy horizon.}{fig:penrose1}

Consider first a point mass inside and an AdS black hole
outside the shell. Within this context we can have both expanding
and collapsing shells, depending upon the direction of the initial
velocity. For the expanding case, the positivity condition, eq.
(\ref{eq:positivity}) leads to two possible final geometries upon
matching to the exterior, as shown in Figs. \ref{fig:penrose1} (c)
and (d). The former contains Cauchy horizons (though not from the
viewpoint of observers on the right half of the diagram).

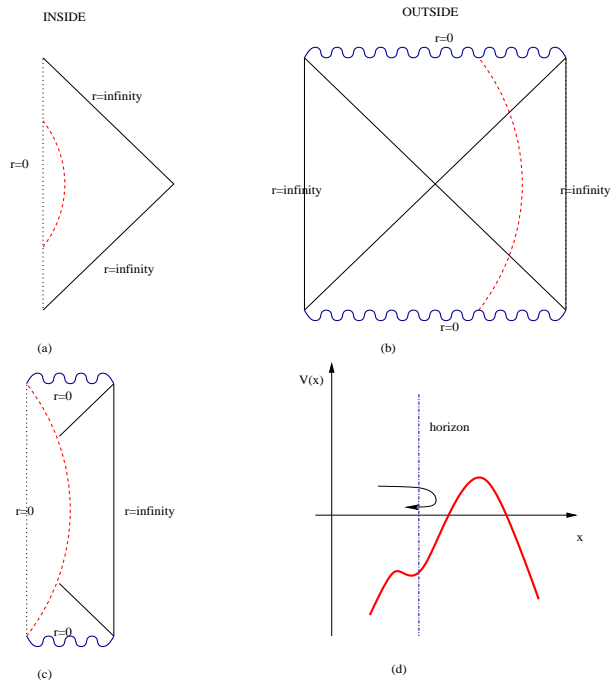
\fig{8cm}{9cm}{penrose2}{\small Sketch of Penrose diagrams of a flat space
inside and a black hole outside the collapsing shell (Figs. (a) and
(b)) and a combined diagram (Fig. (c)). }{fig:penrose2}

The diagrams for a collapsing shell can be obtained by
considering the same procedure in Fig. \ref{fig:penrose2}. Here the shell forms an event horizon
in finite proper time, collapsing into a black hole; the
time-reversed version of this (in which the shell expands out of a
white hole) is also shown in the diagram.

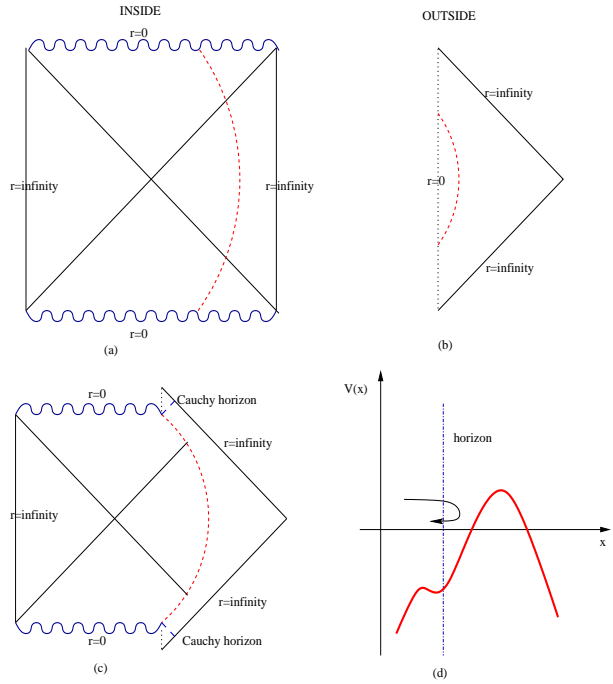
\fig{8cm}{9cm}{penrose3}{\small Sketch of Penrose diagrams of a
black hole inside and a flat space outside the collapsing shell (Figs. (a)
and (b)) and a combined diagram (Fig. (c)); the latter shows that a Cauchy
horizon appears.}{fig:penrose3}

Inverting the exterior and interior, we obtain the
situation depicted in Figs. \ref{fig:penrose3} and
\ref{fig:penrose4} for  contracting and expanding cases
respectively. An interesting feature can be found for the
collapsing shell case (Fig. \ref{fig:penrose3}). In the combined
figure (Fig. \ref{fig:penrose3} (c)), we see that an observer
external to the shell will eventually have the singularity within his/her
past light cone, signalling the appearance of a Cauchy horizon. This can be
avoided for special trajectories of the shell, in which null infinity for the external
observer ends at the endpoint of the shell trajectory in the Penrose diagram.

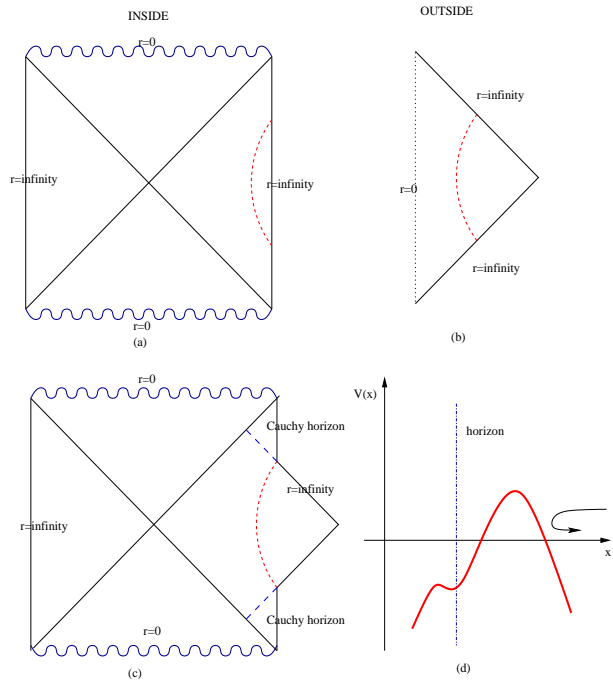
\fig{8cm}{9cm}{penrose4}{\small Sketch of Penrose diagrams of a
black hole inside and a flat space outside an expanding shell (Figs.
(a) and (b)) and the combined diagram (Fig. (c)). Note the resultant
Cauchy horizon in case (c).}{fig:penrose4}

 \fig{8cm}{8.8cm}{penrose5}{\small
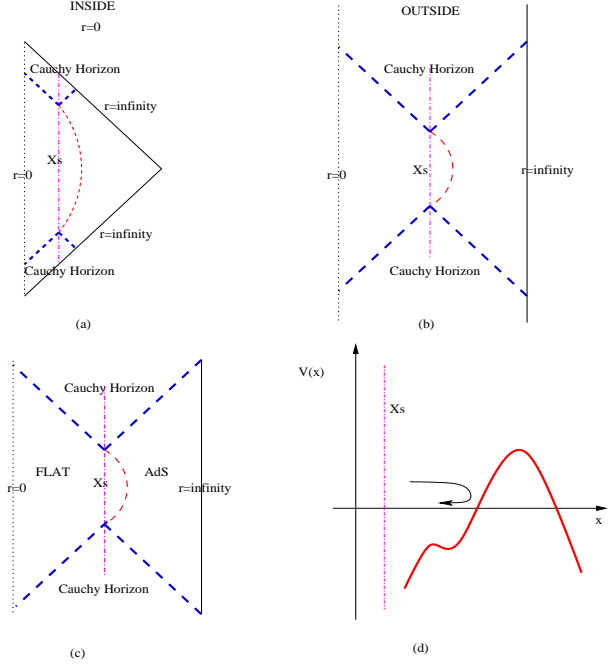
Sketch of Penrose diagrams of a flat space inside and AdS spacetime outside a
collapsing shell with pressure (Figs. (a) and (b)) and a combined
diagram (Fig. (c)). In this case there is a Cauchy horizon that forms where
stress-energy tensor diverges at $x=x_{s}$. }{fig:penrose5}

For a shell with pressure the stress-energy tensor diverges
at $x=x_s$ (where the effective potential also diverges), forming a
singular ring and a Cauchy horizon. The relevant diagrams are shown
in Fig. \ref{fig:penrose5} for a collapsing shell, which also yields
a cosmic censorship violation.

\section{Jacobian Elliptic Functions}
\label{sec:appendix}
In this appendix, we shall briefly introduce Jacobian elliptic functions and its properties.
We can define a doubly periodic elliptical function with real parameters, $m$ and $m_{1}$, where $m+m_{1}=1$, as
\begin{equation}
K(m)=K=\int^{\pi/2}_{0} \frac{d\theta}{\sqrt{1-m \sin^2\theta}},~~iK'(m)=iK'=i\int^{\pi/2}_{0}\frac{d\theta}{\sqrt{1-m_{1} \sin^2\theta}},
\end{equation}
where $0\le m \le 1$.
Note that $K$ and $K'$ are real numbers. Here we denote the points, $0$, $K$, $K+iK'$, $iK'$ by $s$, $c$, $d$, $n$ respectively, which are at the vertices of a rectangle and showing a repeated pattern indefinitely. Now the Jacobian elliptic functions can be defined with respect to an ingetral,
\begin{equation}
\label{eq:integ}
u=\int^{\varphi}_{0}\frac{d\theta}{\sqrt{1-m\sin^2\theta}},
\end{equation}
where the angle $\varphi$ is called the amplitude. Then, we define
\begin{eqnarray}
&& {\rm JacobiSN}(u,k) = \sin\varphi,\\
&& {\rm JacobiCN}(u,k) = \cos\varphi,\\
&& {\rm JacobiDN}(u,k) = \sqrt{1-m\sin^2\varphi}.
\end{eqnarray}
Here we simply denote ${\rm JacobiSN}(u,k)$ as $sn(u|m)$, hereafter, where $m=k^2$.
There are some useful relations of the Jacobian functions to the copolar trio, $sn$, $cn$, $dn$, such that
\begin{eqnarray}
&& cd(u|m)=\frac{cn(u|m)}{dn(u|m)},~~dc(u|m)=\frac{dn(u|m)}{cn(u|m)},~~ns(u|m)=\frac{1}{sn(u|m)}\nonumber\\
&& sd(u|m)=\frac{sn(u|m)}{dn(u|m)},~~nc(u|m)=\frac{1}{cn(u|m)},~~ds(u|m)=\frac{dn(u|m)}{sn(u|m)}\nonumber\\
&& nd(u|m)=\frac{1}{dn(u|m)},~~sc(u|m)=\frac{sn(u|m)}{cn(u|m)},~~cs(u|m)=\frac{cn(u|m)}{sn(u|m)}.\nonumber
\end{eqnarray}
If the parameter $m$ is a positive number, there are some useful relations for the negative parameter. Defining new parameters as $\mu = m/(1+m)$, $\mu_{1}=1/(1+m)$, and $v=u/\sqrt{\mu_{1}}$, where $0<\mu<1$, we have
\begin{eqnarray}
&& sn(u|-m) = \sqrt{\mu_{1}}sd(v|\mu),\label{eq:snm}\\
&& cn(u|-m)=cd(v|\mu), \label{eq:cnm}\\
&& dn(u|-m)=nd(v|\mu). \label{eq:dnm}
\end{eqnarray}
The Jacobian elliptic function is a real function for the real parameters and variables. If we  consider the $SN$ function, $SN(au,m)$, where $a$ and $m$ are constants, then the function has the following properties:
\begin{eqnarray}
&& sn(au|m) \in {\mathcal R}~~(a \in {\mathcal R}, m \in {\mathcal R}~{\rm or}~a \in {\mathcal R}, m \in {\mathcal I})\\
&& sn(au|m) \in {\mathcal I}~~(a \in {\mathcal I}, m \in {\mathcal R}~{\rm or}~a \in {\mathcal I}, m\in {\mathcal I}).
\end{eqnarray}
There are Jacobi's imaginary transformations for the imaginary values of parameters,
\begin{equation}
sn(iu|m) = isc(u|m_{1}),~~cn(iu|m)=nc(u|m_{1}),~~dn(iu|m)=dc(u|m_{1}),\label{eq:jit}
\end{equation}
which are useful to convert a function to a simple form. Moreover, there are some useful relations for real parameters, called Jacobi's real transformation. For $m>0$, defining $\mu=1/m$ and $v=\sqrt{m}u$, then we have
\begin{eqnarray}
\label{eq:reciprocal}
&& sn(u|m) = \sqrt{\mu} sn(v|\mu), \label{eq:jrtsn} \\
&& cn(u|m) = dn(v|\mu), \label{eq:jrtcn}\\
&& dn(u|m) = cn(v|\mu). \label{eq:jrtdn}
\end{eqnarray}
If $m>1$, then $m^{-1}=\mu<1$, which implies that real parameters of elliptic functions always lies between $0$ and $1$.
More details on the further properties on the Jacobian elliptic functions are shown in the ref. \cite{as}.


\begin{thebibliography}{99}
\bibitem{btz} M. Ba$\tilde{\rm n}$ados, C. Teitelboim, and J. Zanelli,
  Phys. Rev. Lett. {\bf 69}, 1849 (1992).
\bibitem{3dgravity} M. Ba$\tilde{\rm n}$ados, M. Henneaux,
  C. Teitelboim, and J. Zanelli, Phys. Rev. D{\bf 48}, 1506 (1993);
  G. T. Horowitz and D. L. Welch, Phys. Rev. Lett. {\bf 71}, 328
  (1993); N. Kaloper, Phys. Rev. D{\bf 48}, 2598 (1993); S. Carlip and
  C. Teitelboim, Phys. Rev. D{\bf 51}, 622 (1995); S. Carlip,
  Phys. Rev. D{\bf 51}, 632 (1995); O. Coussaert, M. Henneaux, and
  P. van Briel, Class. Quantum Grav. {\bf 12}, 2961 (1995); S. Carlip,
  Class. Quantum Grav. {\bf 12}, 2853 (1995); S.-W. Kim, W. T. Kim,
  Y.-J. Park, and H. Shin, Phys. Lett. B{\bf 392}, 311 (1997); S. Carlip,
  Phys. Rev. D{\bf 55}, 878 (1997); M. Ba$\tilde{\rm n}$ados and
  A. Gomberoff, Phys. Rev. D{\bf 55}, 6162 (1997); D. Birmingham,
  I. Sachs, and S. Sen, Phys. Lett. B{\bf 424}, 275 (1998); S. Carlip,
  Class. Quantum Grav. {\bf 15}, 3609 (1998); S. Hyun, W. T. Kim, and
  J. Lee, Phys. Rev. D{\bf 59}, 084020 (1999); M. Ba$\tilde{\rm
  n}$ados, AIP Conf. Proc. {\bf 490}, 198 (1999) (and references not
  listed here therein).
\bibitem{mr}  R. B. Mann and S. F. Ross, Phys. Rev. D{\bf 47}, 3319 (1993).
\bibitem{gak} S. Giddings, J. Abbott, and K. Kuchar,
  Gen. Relativ. Gravit. {\bf 16}, 751 (1984).
\bibitem{mat}  H. J. Matschull, Class. Quantum Grav. {\bf 16}, 1069
  (1999).
\bibitem{3dcollapse} F. Pretorius and M. W. Choptuik, Phys. Rev. D{\bf
    62}, 124012 (2000); V. Husain and M. Olivier, Class. Quantum
    Grav. {\bf 18}, L1 (2001); D. Garfinkle, Phys. Rev. D{\bf 63} 044007
    (2001); G. Clement and A. Fabbri, Nucl. Phys. B{\bf 630}, 269
    (2002); E. W. Hirschmann, A. Wang, and Y. Wu, Class. Quantum
    Grav. {\bf 21}, 1791 (2004); S. Gutti, Class. Quantum Grav. {\bf 22}, 3223
  (2005).
  \bibitem{mr2} R. B. Mann and S. F. Ross, Class. Quantum Grav. {\bf 9},
  2335 (1992).

\bibitem{kmp} A. Kamenshchik, U. Moschella, and V. Pasquier, Phys. Lett.%
\textbf{B511}, 265 (2001).

\bibitem{btv} N. Bilic, G. B. Tupper, and R. D. Viollier,  Phys. Lett. B
\textbf{535}, 17 (2002).

\bibitem{bbs} M. C. Bento, O. Bertolami, and A. A. Sen,  Phys. Rev. D \textbf{66}%
, 043507 (2002).

\bibitem{chap} S. Chaplygin, Sci. Mem. Moscow Univ. Math. Phys. \textbf{21}%
, 1 (1904).

\bibitem{tsien} H.-S. Tsien, J. Aeron. Sci. \textbf{6}, 399 (1939).

\bibitem{karman} T. von Karman, J. Aeron. Sci. \textbf{8}, 337 (1941).

\bibitem{bh} M. Bordemann and J. Hoppe, Phys. Lett. \textbf{B317}, 315
(1993).

\bibitem{hjp} J. Hoppe, hep-th/9311059; R. Jackiw and  A. P. Polychronakos,
Phys. Rev. \textbf{D62}, 085019 (2000); R. Jackiw, \textit{A
Particle Field Theorist's Lectures on Supersymmetric,  Non-Abelian
Fluid Mechanics and d-Branes}, \textrm{[arxiv:physics/{0010042}]}.

\bibitem{bj} D. Bazeia and R. Jackiw, Ann. Phys. (N.Y.) \textbf{270}, 246
(1998).

\bibitem{coas} J. C. Fabris, S. V. B. Goncalves, and P. E. de Souza,  Gen.
Rel. Grav. \textbf{34}, 2111 (2002); H. B. Benaoum, \textrm{hep-th/{0205140}}%
; J. C. Fabris,  S. V. B. Goncalves, and P. E. de Souza, \textrm{astro-ph/{%
0207430}}; A. Dev,  D. Jain, and J. S. Alcaniz, Phys. Rev. D
\textbf{67}, 023515 (2003); M. Makler,  S. Q. de Oliveira, and L.
Waga, Phys. Lett. B \textbf{555}, 1 (2003); M. C. Bento,  O.
Bertolami, and A. A. Sen, Phys. Rev. D \textbf{67}, 063003 (2003);
J. S. Alcaniz,  D. Jain, and A. Dev, Phys. Rev. D \textbf{67},
043514 (2003); R. Bean and O. Dore,  Phys. Rev. D \textbf{68},
023515 (2003); M. C. Bento, O. Bertolami, and A. A. Sen,  Phys.
Lett. B \textbf{575}, 172 (2003); L. Amendola, F. Finelli, C.
Burigana, and D. Carturan, JCAP \textbf{0307}, 005 {2003}; O.
Bertolami, A. A. Sen,  S. Sen, and P. T. Silva, Mon. Not. Roy.
Astron. Soc. \textbf{353}, 329 (2004); T. Barreiro and  A. A. Sen,
Phys. Rev. D \textbf{70}, 124013 (2004); U. Debnath, A. Banerjee,
and  S. Chakraborty, Class. Quant. Grav. \textbf{21}, 5609 (2004);
A. A. Sen and R. J. Scherrer,  Phys. Rev. D \textbf{72}, 063511
(2005); H.-S. Zhang and Z.-H. Zhu,  Phys. Rev. D \textbf{73}, 043518
(2006); M. C. Bento, O. Bertolami, M. J. Reboucas,  and P. T. Silva,
Phys. Rev. D \textbf{73}, 043504 (2006).
\bibitem{lobo} F. S. N. Lobo, Phys. Rev. D\textbf{73}, 064028 (2006).
\bibitem{dc} U. Debnath and S. Chakraborty, \textrm{[arxiv:gr-qc/{0601049}]}.
\bibitem{isr} W. Israel, Nuovo Cimento B \textbf{44}, 1 (1966).
\bibitem{chase} J. E. Chase, Nuovo Cimento B \textbf{67}, 136 (1970).
\bibitem{poi} E. Poisson, {\it A Relativist's Toolkit: The Mathematics
    of Black-Hole Mechanics}, Cambridge University Press, (2004).
  \bibitem{Olea}J. Crisostomo and R. Olea, Phys.Rev. {\bf D69},  104023
  (2004).
  \bibitem{Steif} Y. Peleg and A. Steif, Phys.Rev. {\bf D51}, 3992 (1995).
  \bibitem{Alberghi}G.L. Alberghi, D.A. Lowe and M. Trodden, JHEP {\bf 9907} 020
  (1999).
\bibitem{MyHub} B. Freivogel, V. E. Hubeny, A. Maloney, R. Myers,
M. Rangamani, and  S. Shenker, JHEP {\bf 0603}, 007 (2006).
\bibitem{cf} N. J. Cornish and N. E. Frankel, Phys. Rev. D\textbf{43}, 2555 (1991).
\bibitem{scalarcollapse}F. Pretorius and M. Choptuik, Phys. Rev. D 62, 124012
(2000); V. Husain and M. Olivier, Class. Quantum Grav. 18, L1
(2001); D. Garfinkle, Phys. Rev. D 63, 044007 (2001).
\bibitem{hdfluid}R. Goswami and P. Joshi, \textrm{[arxiv:gr-qc/0608136]}.
\bibitem{as} M. Abramowitz and I. A. Stegun, {\it Handbook of
              Mathematical Functions} (Dover Publications Inc., New
              York, ninth printing, 1970).
\end{thebibliography}
\end{document}